\documentclass[epj]{svjour}
\usepackage[latin1]{inputenc}
\usepackage{graphicx}
\usepackage{color}

\usepackage{citesort}
\usepackage{amssymb}
\usepackage{amsmath}
\usepackage{xspace}
\usepackage{bm}

\newfont{\tensy}{cmsy10}

\newcommand{\ie}[0]{i.e.\@\xspace}
\newcommand{\eg}[0]{e.g.\@\xspace}

\newcommand{\UP}[0]{\uparrow}
\newcommand{\DO}[0]{\downarrow}

\newcommand{\on}{\hat{n}}

\newcommand{\oQ}{\hat{Q}}

\newcommand{\oh}{\mbox{$\frac{1}{2}$}}

\newcommand{\om}[0]{\omega}

\newcommand{\EF}{E_\text{F}}
\newcommand{\kF}{k_\text{F}}

\newcommand{\nag}{{\phantom{\dag}}}

\newcommand{\las}[0]{\langle}
\newcommand{\ras}[0]{\rangle}

\newcommand{\la}[0]{\left\las}
\newcommand{\ra}[0]{\right\ras}
\newcommand{\ket}[1]{\left|#1\ra}
\newcommand{\bra}[1]{\la#1\right|}

\usepackage{pdfpages}
\pdfpageattr {/Group << /S /Transparency /I true /CS /DeviceRGB>>}

\begin{document}
\title{Density waves in strongly correlated quantum chains}
\author{Martin Hohenadler\inst{1} \and Holger Fehske\inst{2}}
\institute{Institut f\"ur Theoretische Physik und Astrophysik, Universit\"at
  W\"urzburg, 97074 W\"urzburg, Germany \and Institut f\"ur
  Theoretische Physik, Ernst-Moritz-Arndt-Universit\"at Greifswald, 17487 Greifswald, Germany}
\date{Received: date / Revised version: date}
%
\abstract{
We review exact numerical results for one-dimensional quantum systems with
half-filled bands. The topics covered include Peierls transitions in Holstein,
Fr\"ohlich, Su-Schrieffer-Heeger, and Heisenberg models with quantum phonons, competing
fermion-boson and fermion-fermion interactions, as well as symmetry-protected
topological states in fermion and anyon models. 
%
} 
\maketitle
%


\section{Introduction}\label{sec:intro}

The properties of quasi-one-dimensional materials such as conjugated
polymers, charge-transfer salts, halogen-brid{\-}ged or organic
superconductors are the result of a subtle interplay of charge, spin, and
lattice fluctuations, in addition to the unique effects of one-dimensional
(1D) correlated quantum systems. This has stimulated intense research efforts
on paradigmatic fermion and fermion-boson models~\cite{TNYS91,BS93}. In
particular, the question how a quasi-1D material evolves
from a metal---either a Tomonaga-Luttin{\-}ger liquid
(TLL)~\cite{To50,Lu63} or a Luther-Emery liquid (LEL) \cite{Lu.Em.74}---to an insulator has remained one of
the most heavily debated issues in solid state physics for decades. Apart from band
structure~\cite{Bl29,Wi31} and disorder effects~\cite{An58},
electron-electron and electron-phonon interactions are the driving forces
behind the metal-insulator transition in the majority of cases. Coulomb
repulsion drives the transition to a Mott insulator (MI) with
dominant spin-density-wave (SDW) fluctuations~\cite{Mo90}, whereas the coupling 
to the vibrational modes of the crystal triggers the Peierls transition~\cite{Pe55} to a
long-range ordered charge-density-wave (CDW) or bond-order-wave (BOW)
insulator \cite{Pouget2016332}. If more than one type of interaction is relevant, 
quantum phase transitions (QPTs) between different insulating phases become possible.
Quite generally, retarded boson-mediated interactions are significantly more
difficult to describe theoretically than the instantaneous Coulomb repulsion.

More recently, QPTs between topologically trivial
and nontrivial states have come into the focus of attention~\cite{Mo10,HK10}.
Topological phases possess characteristic zero-energy edge excitations that
reflect the topological features of the bulk~\cite{Ha82b} and may either
arise from topological band structures or from interactions
\cite{0953-8984-25-14-143201}. The topological properties are
protected by certain symmetries (e.g., inversion, time-reversal or dihedral
symmetry~\cite{GW09,PBTO12}).  Such symmetry-protected topological (SPT)
states have short-range quantum entanglement~\cite{CGZLW13} and may
displace more conventional CDW, BOW, or SDW phases. Examples include
dimerized Su-Schrieffer-Heeger (SSH) models \cite{PhysRevLett.89.077002} and the
Haldane insulator~\cite{Hal83}. 

While the basic mechanisms underlying metal--insulator and insulator--insulator
QPTs are well known, their detailed understanding in microscopic models remains a challenging and active field of
research. Convincing evidence for TLL--insulator QPTs has been obtained for the $t$-$V$
model \cite{Giamarchi}, the spinless Holstein and SSH models
(Secs.~\ref{sec:holstein} and~\ref{sec:ssh}), as well as the
Edwards fermion-boson model (Sec.~\ref{sec:edwards}). Minimal settings for
LEL--insulator QPTs are the spinful Holstein and the extended Hubbard
model (Secs.~\ref{sec:holstein} and~\ref{sec:extendedhubbard}). Insulator--metal--insulator or direct
insulator--insulator QPTs have been explored in the extended Hubbard
model (Sec.~\ref{sec:extendedhubbard}), the Holstein-Hubbard model
(Sec.~\ref{sec:holsteinhubbard}), the Holstein-SSH model with competing bond
and site couplings (Sec.~\ref{sec:holsteinssh}), and the SSH model with
additional Cou{\-}lomb interaction (Sec.~\ref{sec:sshuv}). Extended
Falicov-Kimball models (Sec.~\ref{sec:efkm}) exhibit QPTs between semimetals or
semiconductors and excitonic insulators. Extended Hubbard models with
either an additional alternating ferromagnetic spin interaction or a bond
dimerization have topologically trivial density-wave (DW) states but also
SPT phases (Sec.~\ref{sec:ehmdimer}).
Finally, more exotic bosonic or even anyonic models that can be realized, in
particular, with highly tunable cold atoms in optical lattices \cite{BDZ08} exhibit
superfluid, MI, CDW, and SPT states (Sec.~\ref{sec:eahm}).

In this contribution, we review the physics of a variety of lattice models for
quasi-1D strongly correlated particle systems. Focusing on results from
numerically exact methods such as Lanczos exact diagonalization \cite{CW85,WF08a}, the density matrix
renormalization group (DMRG) \cite{Wh92,Sc05,JF07,McC08}, and continuous-time quantum
Monte Carlo (QMC) \cite{Rubtsov05,Assaad07,Gull_rev,Sandvik02,SandvikPhononsa,CH05}, we discuss
ground-state and spectral properties and relate them to the corresponding 1D
low-energy theories \cite{Voit94,Giamarchi}. Given the enormous literature, we mostly
restrict the scope to half-filled bands for which umklapp scattering can give
rise to QPTs. Section~\ref{sec:fermion-boson} is
devoted to the effects of fermion-boson coupling, whereas Coulomb interaction
will be discussed in Sec.~\ref{sec:coulomb}. In
Sec.~\ref{sec:symmetryprotected} we review recent work on SPT states. Finally, we conclude in Sec.~\ref{sec:conclusions}.

\section{Density waves from fermion-boson coupling}\label{sec:fermion-boson}

\subsection{Holstein and Fr\"ohlich-type models}\label{sec:holstein}

Perhaps the simplest example of a quantum system of coupled fermions and bosons
are charge carriers interacting with lattice vibrations, as described by the Holstein Hamiltonian~\cite{Ho59a} ($\hbar=1$)
\begin{align}\label{eq:holstein}
  \hat{H}_{\rm Hol}
  = -t\sum_{\las i,j\ras\sigma} \hat{c}^\dagger_{i\sigma} \hat{c}^{\phantom{\dagger}}_{j\sigma} 
  +\omega_0\sum_i \hat{b}^{\dagger}_i \hat{b}^{\phantom{\dagger}}_{i}
  -g\omega_0\sum_{i}
  (\hat{b}^{\dagger}_i + \hat{b}^{\phantom{\dagger}}_{i}) \on_{i}\,.     
\end{align}
It accounts for a single tight-binding electron band emerging from
nearest-neighbor hopping, quantum phonons in the
harmonic approximation, and a local density-displace{\-}ment electron-phonon
coupling. Here, $\on_i=\sum_\sigma\on_{i\sigma}$ and
$\on_{i\sigma}=\hat{c}^{\dagger}_{i\sigma} \hat{c}^{}_{i\sigma}$,
where $\hat{c}^{\dagger}_{i\sigma}$ ($\hat{c}^{}_{i\sigma}$) creates (annihilates) a spin-$\sigma$
electron at site $i$ of a 1D lattice with $L$ sites. Similarly,
$\hat{b}^{\dagger}_{i}$ ($\hat{b}_{i}$) creates (annihilates) a dispersionless optical (Einstein)
phonon of frequency  $\omega_0$. Half-filling corresponds to
$n=\las\on_i\ras=1$ ($n=1/2$) for spinful (spinless) fermions.

The physics of the Holstein model is governed by the
competition between the itinerancy of the electrons and the tendency of the
electron-phonon coupling to  ``immobilize'' them. Importantly, the 
interaction is retarded in nature, as described by the adiabaticity ratio
$\omega_0/t$. Throughout this article, we use $t$ as the energy unit. The
electron-phonon coupling is often parameterized by
$\lambda=\varepsilon_\text{p}/2t$ in the adiabatic regime ($\omega_0/t\ll 1$), and by 
$g^2=\varepsilon_\text{p}/\omega_0$ in the anti-adiabatic regime ($\omega_0/t \gg 1$)
\cite{WF98a,WF98b,PhysRevB.56.4484}. For the single-particle case, where $\varepsilon_\text{p}$ is the
polaron binding energy, the Holstein model has provided important insight into the notoriously difficult problem of
polaron formation and self-trapping~\cite{Ho59a,AFT10}. The half-filled case
considered here provides a framework to investigate the even
more intricate problem of the Peierls metal-insulator QPT of
spinless \cite{Hirsch83a,CB84,ZFA89,WF98b,MHM96,BMH98,PhysRevB.71.235118,Hohenadler06,arXiv:1704.07913}
or spinful fermions \cite{Hirsch83a,Guinea83,CB84,JeZhWh99,WFWB00,PhysRevB.67.081102,CH05,HC07,PhysRevB.71.205113,Ba.Bo.07,EF09a,FHJ08,EF10,Barkim2015}.

\begin{figure}
\centering
\includegraphics[width=0.45\textwidth]{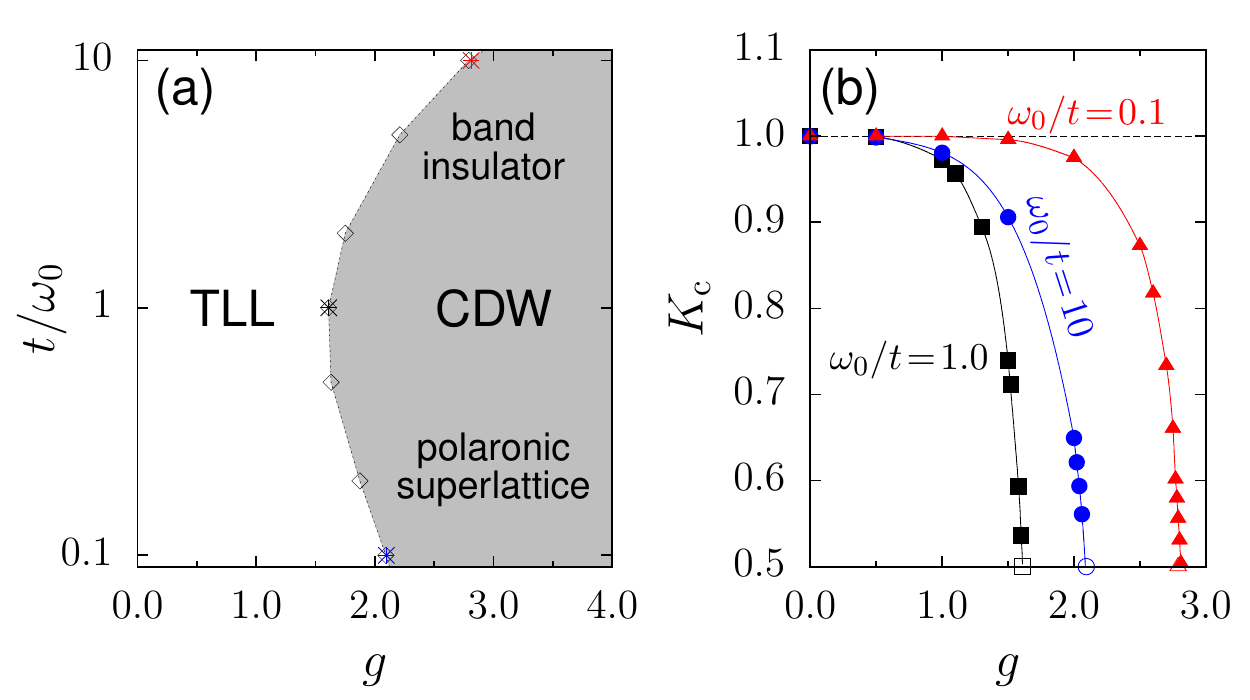}
\caption{\label{fig:holspinless-phasediagram}
(a) Phase diagram and (b) TLL parameter $K_\text{c}$
of the spinless Holstein model from DMRG calculations \cite{EF09a}.}
\end{figure}

The {\it spinless} Holstein model is obtained from
Eq.~(\ref{eq:holstein}) by dropping spin sums and indices.
Figure~\ref{fig:holspinless-phasediagram}(a) shows the
corresponding ground-state phase diagram from fermion-boson
pseudo-site DMRG calculations \cite{JF07}. At a critical coupling
$g_c(\omega_0)$, a QPT from a TLL to a CDW insulator with long-range
$q=2\kF=\pi$ order (alternating occupied and empty sites) and a $2\kF$
Peierls lattice distortion~\cite{Pe55} takes place. The insulating state can be classified as a
traditional band insulator in the adiabatic regime, and as a polaronic superlattice
in the anti-adiabatic regime~\cite{FHW00,Hohenadler06}. Numerical evidence for the Kosterlitz-Thouless
\cite{KT73} transition expected from the low-energy TLL description and
the mapping to the $t$-$V$ model at strong coupling \cite{Hirsch83a} comes
from, \eg, XXZ-model physics for large $\omega_0$ \cite{Hohenadler06} and a
cusp in the fidelity susceptibility \cite{arXiv:1704.07913}.

The TLL charge parameter $K_\text{c}$---determining the decay of correlation
functions \cite{Voit94,Giamarchi}---from a finite-size scaling of the
long-wavelength limit of the charge structure factor
$S_\text{c}(q)=\frac{1}{L}\sum_{j,l} e^{iq(j-l)}\langle \on_j \on_l\rangle$ according
to \cite{Giamarchi,EF09a}
\begin{equation}\label{eq:sc}
K_\text{c} = \lim_{L\to\infty} K_\text{c}(L)\,,\quad K_\text{c}(L)= \pi  \frac{S_\text{c}(q_1)}{q_1}\,,\quad  q_1=\frac{2\pi}{L}
\end{equation}
is shown in Fig.~\ref{fig:holspinless-phasediagram}(b).
Contrary to earlier numerical results~\cite{FHW00,WF98b}, the TLL turns out
to be repulsive ($K_\text{c} < 1$) for any
$\omega_0$~\cite{EF09a}. Accordingly, charge correlations ($\sim r^{-2K_\text{c}}$) dominate over
pairing correlations ($\sim r^{-2/K_\text{c}}$) throughout the TLL phase and show
a crossover from weak to strong $2\kF$ power-law correlations with increasing
coupling \cite{PhysRevB.92.245132,arXiv:1704.07913}. As shown in Fig.~\ref{fig:holspinless-phasediagram}, $K_\text{c}=1/2$ at the critical point,
as expected for the umklapp-driven Mott transition in a spinless TLL \cite{Giamarchi}. 

\begin{figure}[t]
\centering
\includegraphics[width=0.45\textwidth]{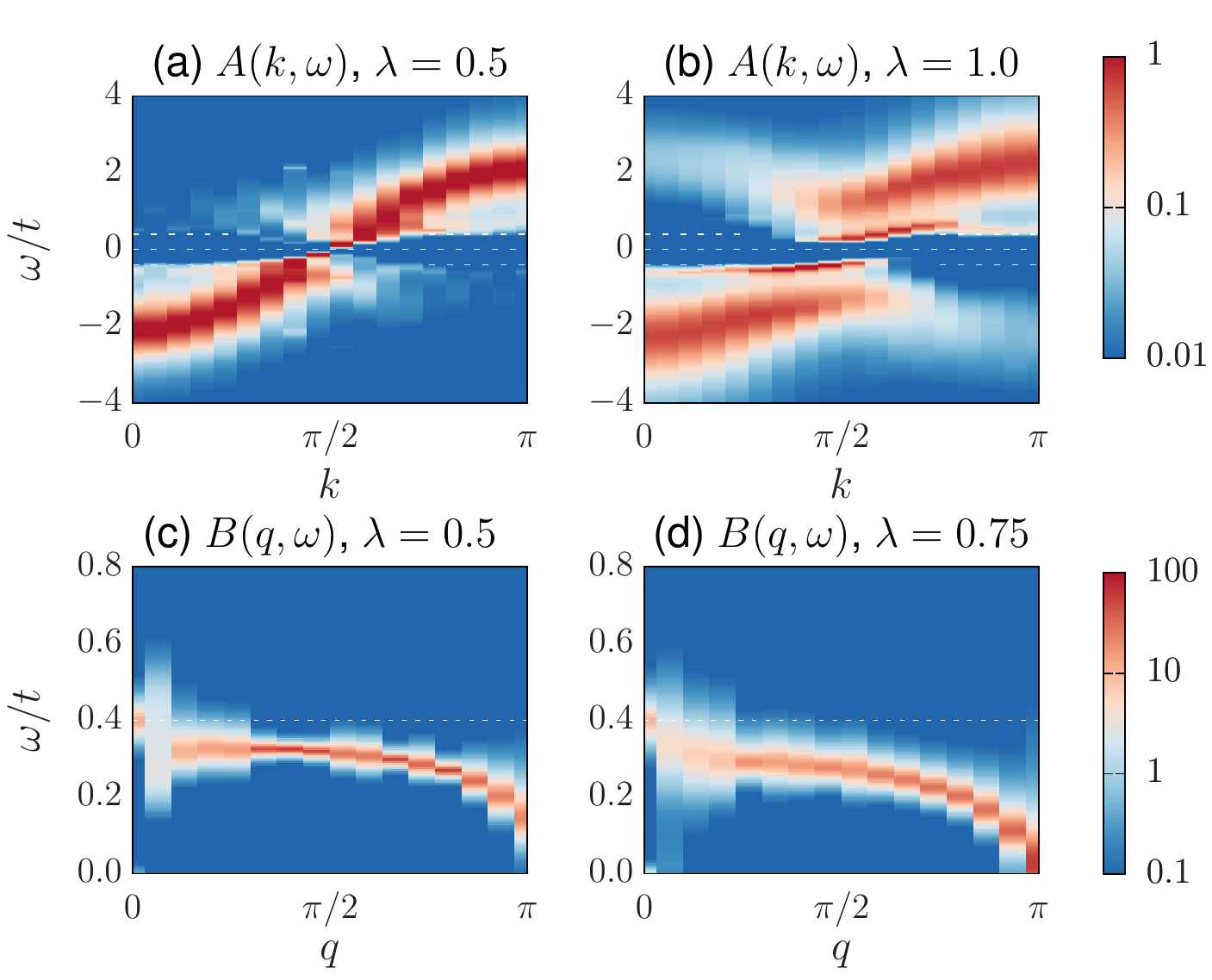}
\caption{\label{fig:holspinless-excitations}
Single-particle [(a),(b)] and phonon [(c),(d)] spectra of the spinless
Holstein model from QMC \cite{Hohenadler10a,PhysRevB.94.245138}. 
Dashed lines indicate $E_\text{F}=0$ and $\om=\pm\om_0$ ($\om_0/t=0.4$).}
\end{figure}

Figure~\ref{fig:holspinless-excitations} shows excitation spectra from
QMC simulations \cite{Hohenadler10a,PhysRevB.94.245138}, namely the 
single-particle spectral function
\begin{align}\label{eq:akw}\nonumber
  A(k,\om)
  =
  \frac{1}{Z}\sum_{mn}
  &
  {|\bra{\psi_m} \hat{c}_{k} \ket{\psi_n}|}^2 (e^{-\beta E_m}+e^{-\beta E_n})
  \\
  &\times
     \delta[\omega - (E_n-E_m)] 
\end{align}
and the phonon spectral function
\begin{align}\nonumber
  B(q,\omega) 
  = 
  \frac{1}{Z} & \frac{1}{\sqrt{2M\omega_0}}
  \sum_{mn} | \langle \psi_m | \hat{b}^\dag_q + \hat{b}^\nag_{-q} | \psi_n\rangle |^2 e^{-\beta E_m}  
  \\
  &\hspace*{5.5em}\times
  \delta[\omega - (E_n-E_m)]\,;
\end{align}
$E_n$ is the eigenvalue for $\ket{\psi_n}$, $Z$ the partition function.

In the adiabatic regime, the single-particle spectrum in the TLL phase
[Fig.~\ref{fig:holspinless-excitations}(a)] is gapless but
significantly modified by the hybridization of charge and phonon modes \cite{Meden94,Hohenadler06,Hohenadler10a}. In the CDW phase,
it exhibits a Peierls gap and backfolded shadow bands
\cite{Vo.Pe.Zw.Be.Ma.Gr.Ho.Gr.00,FeWeHaWeBi03,Hohenadler06,Hohenadler10a} [Fig.~\ref{fig:holspinless-excitations}(b)]. Near the
critical point, soliton excitations \cite{PhysRevLett.42.1698} can
be observed \cite{Hohenadler10a}. The phonon spectrum \cite{PhysRevB.94.245138} reveals the renormalization
of the phonon mode due to electron-phonon coupling. In the adiabatic regime, the mode
softens at the zone boundary in the TLL phase
[Fig.~\ref{fig:holspinless-excitations}(c)], becomes completely soft for
$q=2\kF$ at the critical point [Fig.~\ref{fig:holspinless-excitations}(d)], and hardens
again in the CDW phase \cite{CrSaCa05,SyHuBeWeFe04,Hohenadler10a,PhysRevB.94.245138}.
In contrast, for $\omega_0\gg t$, the phonon mode hardens in the metallic
phase and a central mode appears at $\lambda_c$ \cite{Hohenadler06}. These findings are consistent with
a soft-mode transition for $\omega_0\ll t$ and a central-peak transition for
$\omega_0\gg t$ \cite{Hohenadler06}. 

A complete picture of the physics of the {\em spinful} Holstein
model~(\ref{eq:holstein}) has only emerged recently. Whereas early work
\cite{Hirsch83a,PhysRevB.71.205113,Ba.Bo.07} suggested the absence of a metallic
phase, the existence of the latter has since been confirmed \cite{JeZhWh99,CH05,HC07,FHJ08,EF10,Barkim2015};
for a detailed review see \cite{PhysRevB.92.245132}.  In terms of g-ology \cite{Giamarchi}, the attractive umklapp scattering
arising from the Holstein coupling remains irrelevant for $\lambda<\lambda_c(\omega_0)$. However, for any $\lambda>0$, attractive backscattering
opens a spin gap \cite{PhysRevB87.075149}. Therefore, the metallic phase is in fact a 1D spin-gapped
metal---also known as an LEL \cite{Lu.Em.74}. Using the notation C$x$S$y$
for a system with $x$ ($y$) gapless charge (spin) modes
\cite{PhysRevB.53.12133}, the LEL has C1S0.
For $\lambda>\lambda_c$, umklapp
scattering is relevant and the ground state is a $2\kF$ CDW
insulator (alternating doubly occupied and empty sites) with C0S0. Estimates
for $\lambda_c$ are contained in the phase diagram of the
Holstein-Hubbard model in Fig.~\ref{fig:holhub_phasediagram} in the limit $U=0$.

\begin{figure}[t]
\centering
\includegraphics[width=0.45\textwidth]{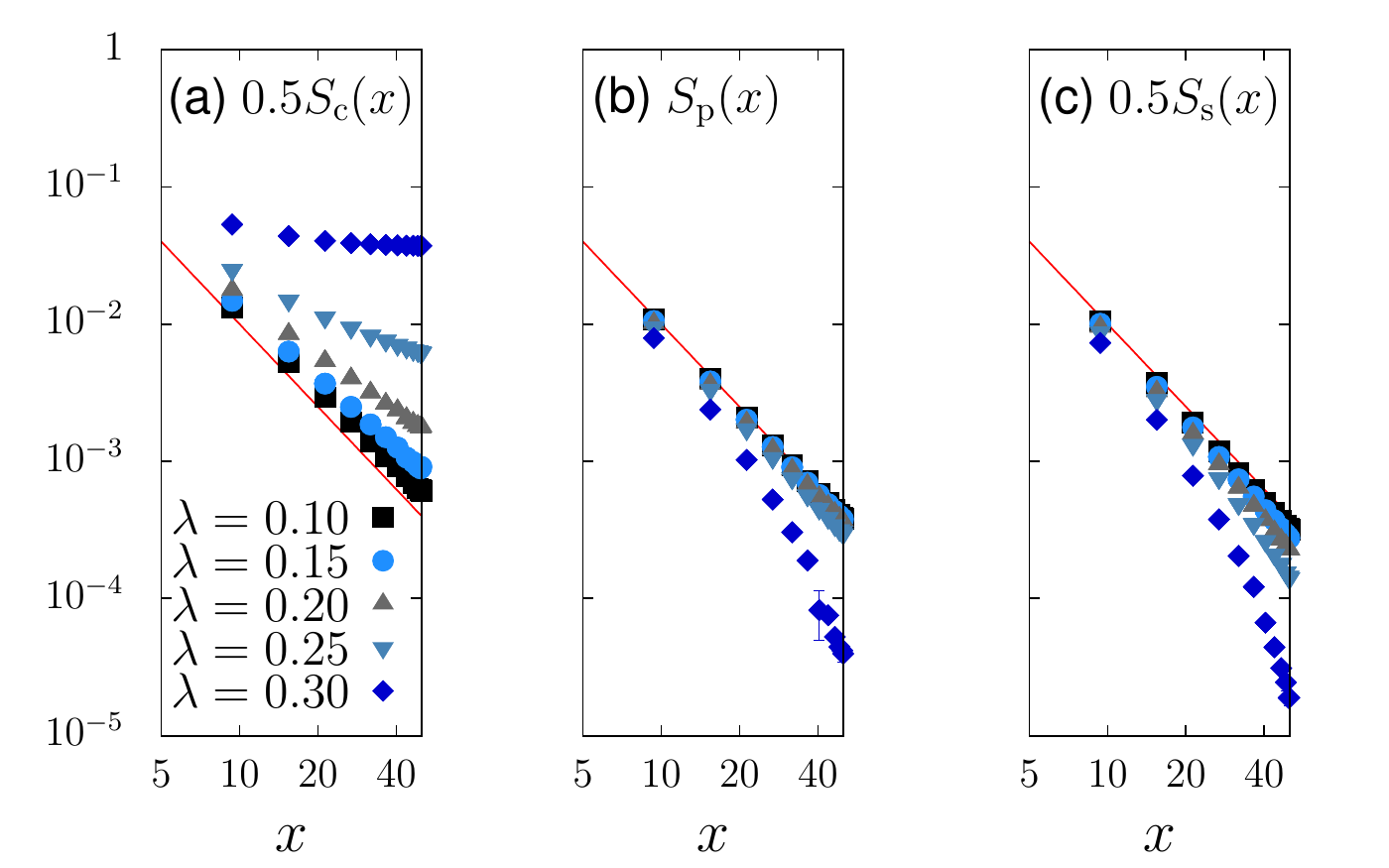}
\caption{\label{fig:holhub_correlators}
Real-space correlation functions of the spinful Holstein model~(\ref{eq:holstein}) 
for (a) charge, (b) pairing, and (c) spin from QMC simulations
\cite{PhysRevB.92.245132}. Here, $\omega_0/t=0.5$,
$x= L\sin \left(\pi r/L\right)$ is the
conformal distance \cite{Cardybook}, and the solid line indicates $1/x^2$.}
\end{figure}

LEL physics and the Peierls QPT are also revealed by the real-space
correlation functions
\begin{align}
  S_\text{c}(r)   &=  \las (\on_r -n) (\on_0-n) \ras\,, \\\nonumber
  S_\text{s}(r) &= \las \hat{S}^x_r \hat{S}^x_0 \ras\,,\\\nonumber
  S_\text{p}(r)       &= \las \hat{\Delta}^\dag_r \hat{\Delta}^\nag_0 \ras \quad
  (\hat{\Delta}_r = \hat{c}^\dag_{r\UP} \hat{c}^\dag_{r\DO})\,,
\end{align}
measuring charge, spin, and s-wave pairing correlations. As in the spinless case, charge correlations
dominate over pairing in the metallic phase
\cite{PhysRevB.84.165123,PhysRevB87.075149,PhysRevB.92.245132}, see Fig.~\ref{fig:holhub_correlators}.
Such behavior necessarily requires a spin gap \cite{Voit98} ($K_\text{s}=0$) and repulsive interactions ($K_\text{c}<1$).
The spin gap complicates the determination of $K_\text{c}$ \cite{PhysRevB.92.245132} but
the correlation functions in Fig.~\ref{fig:holhub_correlators} clearly rule out
claims of dominant pairing \cite{CH05}. Spectral properties of the spinful
Holstein model have also been calculated
\cite{ZhJeWh99,NiZhWuLi05,Ho.As.Fe.12,PhysRevB87.075149,WeAsHo15I}. Most
notably, the single-particle spectrum is gapped even in the metallic phase
(although the spin gap---not taken into account in \cite{Meden94}---is
difficult to detect numerically at weak coupling), and 
the phonon spectrum reveals a soft-mode transition similar to the spinless
case for $\om_0/t<1$ \cite{WeAsHo15I}. 

\begin{figure}
\centering
\includegraphics[width=0.2\textwidth]{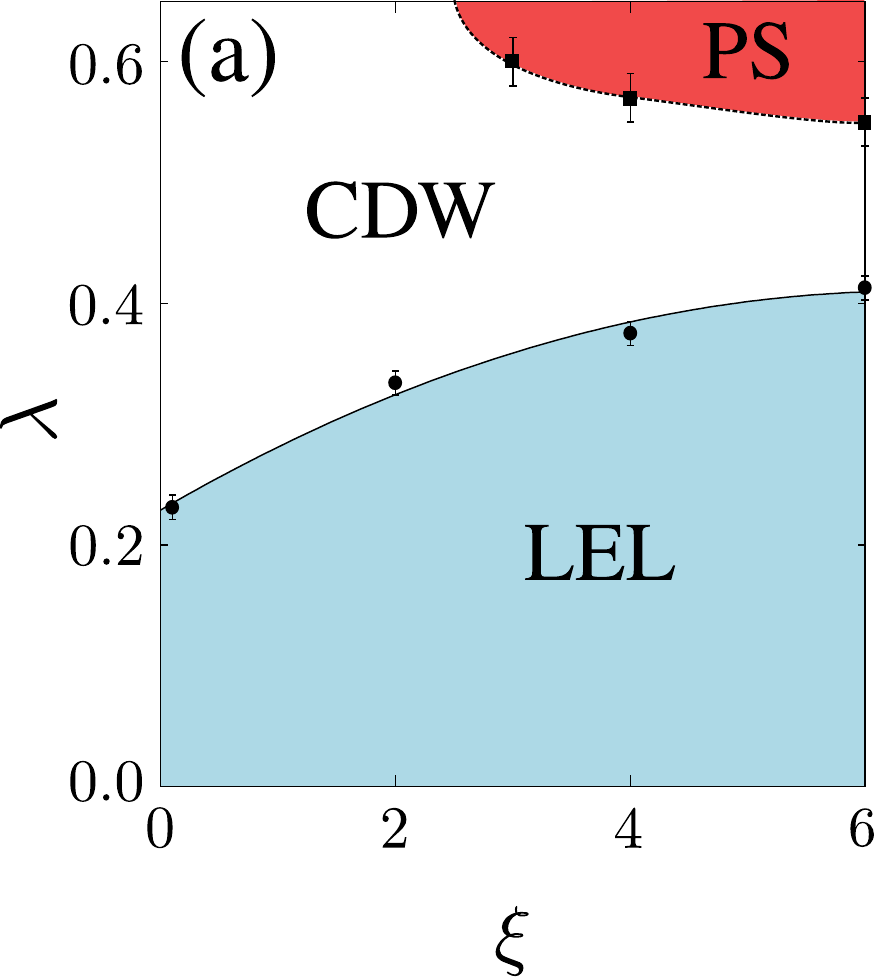}
\includegraphics[width=0.225\textwidth]{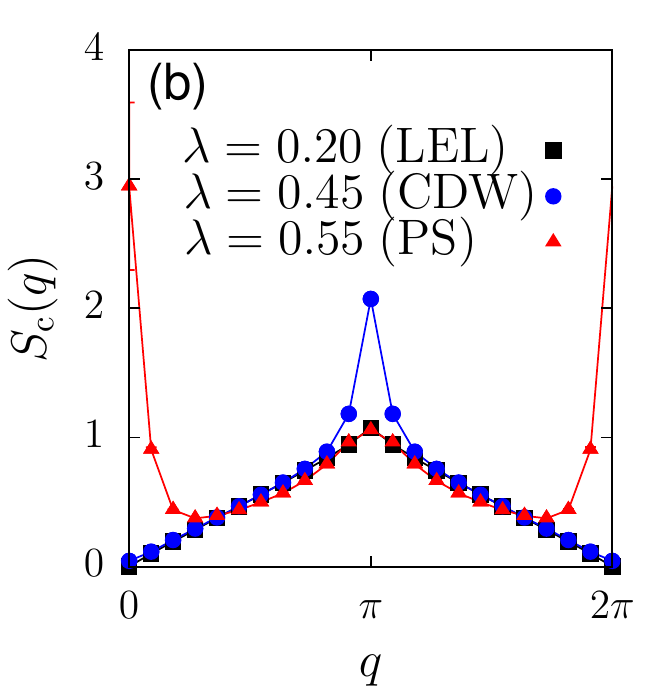}
\caption{\label{fig:froehlich}
(a) Phase diagram and (b) charge structure factor in the three phases
of the nonlocal electron-phonon model~(\ref{eq:froehlich})
from QMC \cite{Ho.As.Fe.12}. Here, $\omega_0/t=0.5$.}
\end{figure}

A local electron-phonon interaction as in the Holstein model~(\ref{eq:holstein}) is {\it a priori} not
justified for materials with incomplete screening. Results for nonlocal interactions in the empty-band limit reveal
significantly reduced polaron and bipolaron masses \cite{0034-4885-72-6-066501}.
More recently, numerical results for half-filling were obtained \cite{Ho.As.Fe.12}. As a function of the
screening length $\xi$, the Hamiltonian \cite{Ho.As.Fe.12} 
\begin{align}\label{eq:froehlich}
  \hat{H}
  = &-t\sum_{\las i,j\ras\sigma} \hat{c}^\dagger_{i\sigma} \hat{c}^{\phantom{\dagger}}_{j\sigma} 
      +\omega_0\sum_i \hat{b}^{\dagger}_i \hat{b}^{\phantom{\dagger}}_{i}
  \\\nonumber
    &- g\omega_0\sum_{ir\sigma} \frac{e^{-r/\xi}}{(r^2+1)^{3/2}}\, (\hat{b}^\dag_{i+r}+\hat{b}^\nag_{i+r})  \on_{i\sigma}
\end{align}
interpolates between a local Holstein and a long-range Fröhlich-type coupling
\cite{AlKo99}. As shown in Fig.~\ref{fig:froehlich}(a), for small to
intermediate $\xi$, the same LEL and CDW phases are found, but $\lambda_c$ is
enhanced with increasing $\xi$. For large
$\xi$ and strong coupling, the nonlocal interaction gives rise to multipolaron
droplets and phase separation (PS) \cite{Ho.As.Fe.12}, as detected from
the $q=0$ divergence of the charge structure factor
[Fig.~\ref{fig:froehlich}(b)] that implies $K_\text{c}=\infty$ and hence a divergent
compressibility \cite{Giamarchi}.  The CDW--PS QPT appears to be of first
order \cite{Ho.As.Fe.12}. Increasing the interaction range at a fixed $\lambda$ drives a CDW--LEL QPT. The
concomitant suppression of CDW order gives rise to degenerate  
pairing and charge correlations in the Fröhlich limit $\xi\to\infty$.

\subsection{Su-Schrieffer-Heeger model}\label{sec:ssh}

\begin{figure}
\centering
\includegraphics[width=0.45\textwidth]{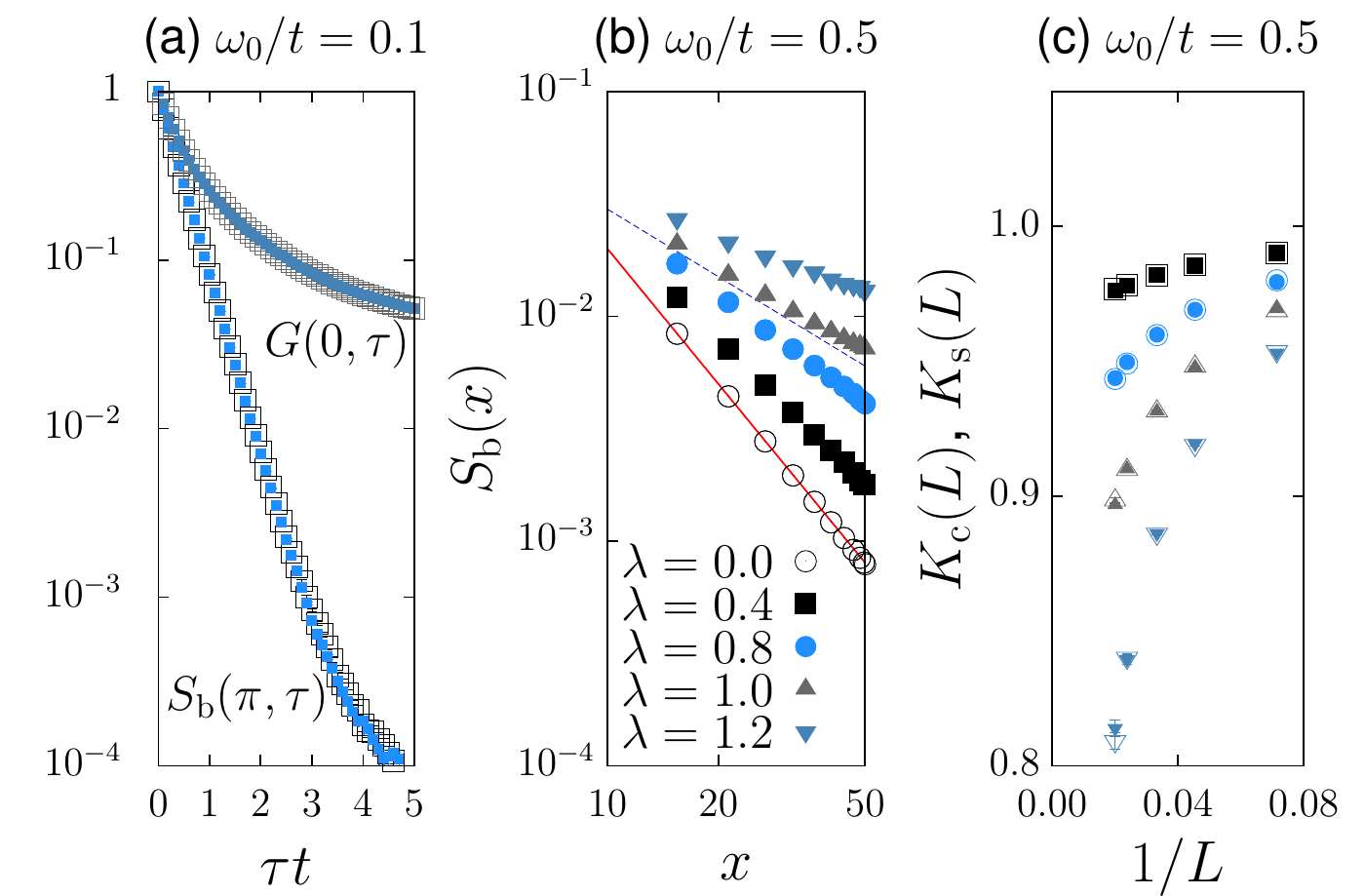}
\caption{\label{fig:ssh}
(a) Single-particle Green function and dynamic bond
structure factor for the spinful optical SSH model~(\ref{eq:ssh}) and the original
SSH model. (b) Real-space bond correlations of
the spinless optical SSH model. The full (dashed) line corresponds to $1/x^2$
($1/x$). (c)
Finite-size estimates of the TLL parameters $K_\text{c}$ and $K_\text{s}$. All results
are from QMC simulations \cite{PhysRevB.91.245147}.
}
\end{figure}

The SSH model of polyacetylene captures fluctuations of the carbon-carbon
bond lengths and their effect on the electronic hopping integral \cite{PhysRevLett.42.1698} (for related
earlier work see \cite{PhysRevLett.25.919}). It has a coupling term of the form
$\sum_{i} \hat{B}_{i} (\oQ_{i+1}-\oQ_i)$, where $\oQ_{i}\sim
\hat{b}^\dag_i+\hat{b}^\nag_i$ is the displacement of atom $i$ from its equilibrium
position and $\hat{B}_i=\sum_\sigma
(\hat{c}^\dag_{i\sigma}\hat{c}^\nag_{i+1\sigma}+\text{H.c.})$. The phonons have an acoustic dispersion $\omega_q=\om_0 \sin (q/2)$. Reviews in the
context of conjugated polymers were given in \cite{Baeriswylreview92,Barfordbook}.
Theoretical arguments \cite{PhysRevB.27.1680,PhysRevB.91.245147} and exact
numerical results \cite{PhysRevB.91.245147} suggest that at half-filling
the SSH model is equivalent to the simpler {\em optical} SSH model \cite{PhysRevB.67.245103}
\begin{equation}\label{eq:ssh}
  \hat{H}_\text{SSH}
  =
  -t \sum_{i} \hat{B}_i +
  \omega_0\sum_i \hat{b}^\dag_i \hat{b}^\nag_i
  -
  g \omega_0
  \sum_{i} 
  \hat{B}_{i}
  (\hat{b}^\dag_i + \hat{b}^\nag_i)
  \,.
\end{equation}
Here, $\hat{b}^\dag_i$ and $\hat{b}^\nag_i$ are associated with an optical phonon mode
describing fluctuations of the bond lengths. Figure~\ref{fig:ssh}
illustrates the quantitative agreement of the two models for the
single-particle Green function and the dynamic bond structure factor, which
can be attributed to the inherent dominance of $q=2\kF=\pi$ order at
half-filling \cite{PhysRevB.91.245147}. 

For spinless fermions, the model~(\ref{eq:ssh}) has a repulsive TLL ground
state with dominant BOW correlations below a critical coupling $\lambda_c$
(for SSH models $\lambda=g^2\omega_0/2t$), and an insulating Peierls ground
state with a long-range $2\kF$ BOW (alternating weak and strong bonds) for
$\lambda>\lambda_c(\omega_0)$
\cite{PhysRevB.27.1680,Ba.Bo.07,PhysRevB.91.245147}. The transition from
power-law to long-range BOW correlations can be seen in Fig.~\ref{fig:ssh}(b)
from
$S_\text{b}(r) = \las (\hat{B}_{r}-\las \hat{B}_{r}\ras)(\hat{B}_{0}-\las
\hat{B}_{0}\ras)\ras$.
At the critical point, the correlation functions are consistent with
$K_\text{c}=1/2$.  In the adiabatic regime, the phase transition is again of the
soft-mode type \cite{PhysRevB.91.245147}.  Apart from the interchange of the
roles of charge and bond degrees of freedom, the spinless SSH model is in
many respects similar to the spinless Holstein model, including spectral and
thermodynamic properties
\cite{PhysRevB.91.245147,PhysRevB.94.155150}. However, subtle differences
arise due to the different symmetries of the two models (class BDI of the
general classification \cite{PhysRevB.78.195125} for the SSH model, class AI
with broken particle-hole and chiral symmetry for the Holstein model)
\cite{PhysRevB.94.155150}. Note that the name SSH model is often used to
refer to the mean-field approximation of the true SSH model, i.e., a
fermionic Hamiltonian with dimerized hopping but no phonons (see also
Sec.~\ref{sec:ehmdimer}).

In contrast to the spinful Holstein model, the spinful SSH model does not have a metallic
phase. Although quantum fluctuations significantly reduce the dimerization
compared to the mean-field solution \cite{PhysRevB.25.7789}, the ground state is an insulating
BOW-Peierls state (C0S0) for any $\lambda>0$ irrespective of $\omega_0$
\cite{Su1982497,PhysRevB.27.1680,PhysRevLett.60.2089,PhysRevB.67.245103,PhysRevB.73.045106,Ba.Bo.07,Barkim2015,PhysRevB.91.245147}.
Direct numerical evidence for this conclusion is shown in Fig.~\ref{fig:ssh}(c). Because
Eq.~(\ref{eq:ssh}) is symmetric under the transformation
$\hat{c}^\nag_{i\DO}\mapsto(-1)^i\hat{c}^\dag_{i\DO}$ that interchanges spin and charge
operators, spin and charge correlators are exactly equal. Therefore, the
finite-size estimate of $K_\text{s}(L)<1$ (a reliable indicator of a spin gap
for models with SU(2) spin symmetry \cite{HC07}) implies
$K_\text{s}=K_\text{c}=0$ by symmetry and hence an insulating state \cite{PhysRevB.91.245147}. The spinful
SSH model hence provides an example where Peierls' theorem \cite{Pe55} holds even for
quantum phonons. This property can be traced back to the fact that
forward scattering vanishes whereas umklapp scattering is repulsive (rather than
attractive, as in the Holstein model) and hence always relevant \cite{Barkim2015}. 
Excitation spectra for the
spinful SSH model closely resemble those of the spinless model in the ordered
phase \cite{PhysRevB.91.245147}.

\subsection{Holstein-Su-Schrieffer-Heeger model}\label{sec:holsteinssh}

Whereas Holstein and SSH models have been studied intensely,
the even more complex problem of competing site and bond couplings---which in principle coexist in most materials---has been
addressed by QMC only recently \cite{PhysRevLett.117.206404} using the Holstein-SSH Hamiltonian
\begin{align}\label{eq:sshhol}\nonumber
  \hat{H}
  =
  &-t \sum_{i} \hat{B}_i +
    \sum_{i\alpha} \omega_{0,\alpha} \hat{b}^\dag_{i,\alpha} \hat{b}^\nag_{i,\alpha} 
    -
    g_\text{s} \omega_{0,\text{s}} 
    \sum_{i} \on_{i} (\hat{b}^\dag_{i,\text{s}}+ \hat{b}^\nag_{i,\text{s}})
  \\
  &-
    g_\text{b}\omega_{0,\text{b}}
    \sum_{i} 
    \hat{B}_{i} (\hat{b}^\dag_{i,\text{b}}+ \hat{b}^\nag_{i,\text{b}})
\end{align}
with independent site ($\alpha=\text{s}$) and bond
($\alpha=\text{b}$) phonon modes as well as corresponding coupling constants $\lambda_\alpha$.

\begin{figure}[t]
\centering
\includegraphics[width=0.375\textwidth]{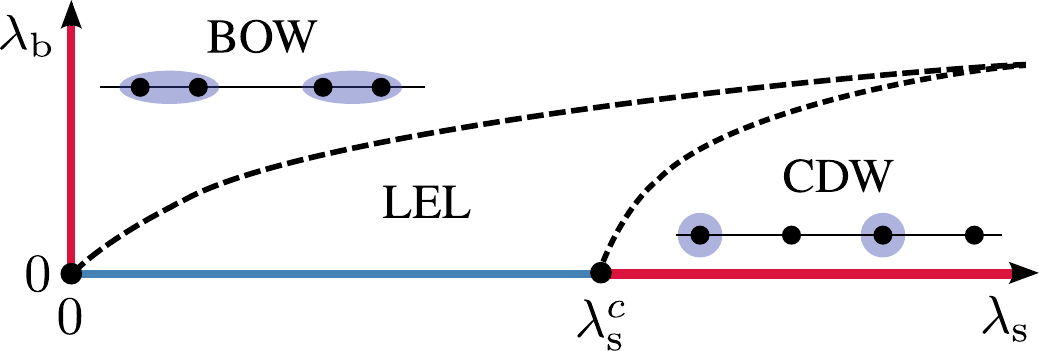}
\caption{\label{fig:sshhol}
Schematic phase diagram of the Holstein-SSH model~(\ref{eq:sshhol}) based on QMC simulations for
$\omega_0/t=0.5$ \cite{PhysRevLett.117.206404}.}
\end{figure}

\begin{figure}[b]
\centering
\includegraphics[width=0.45\textwidth]{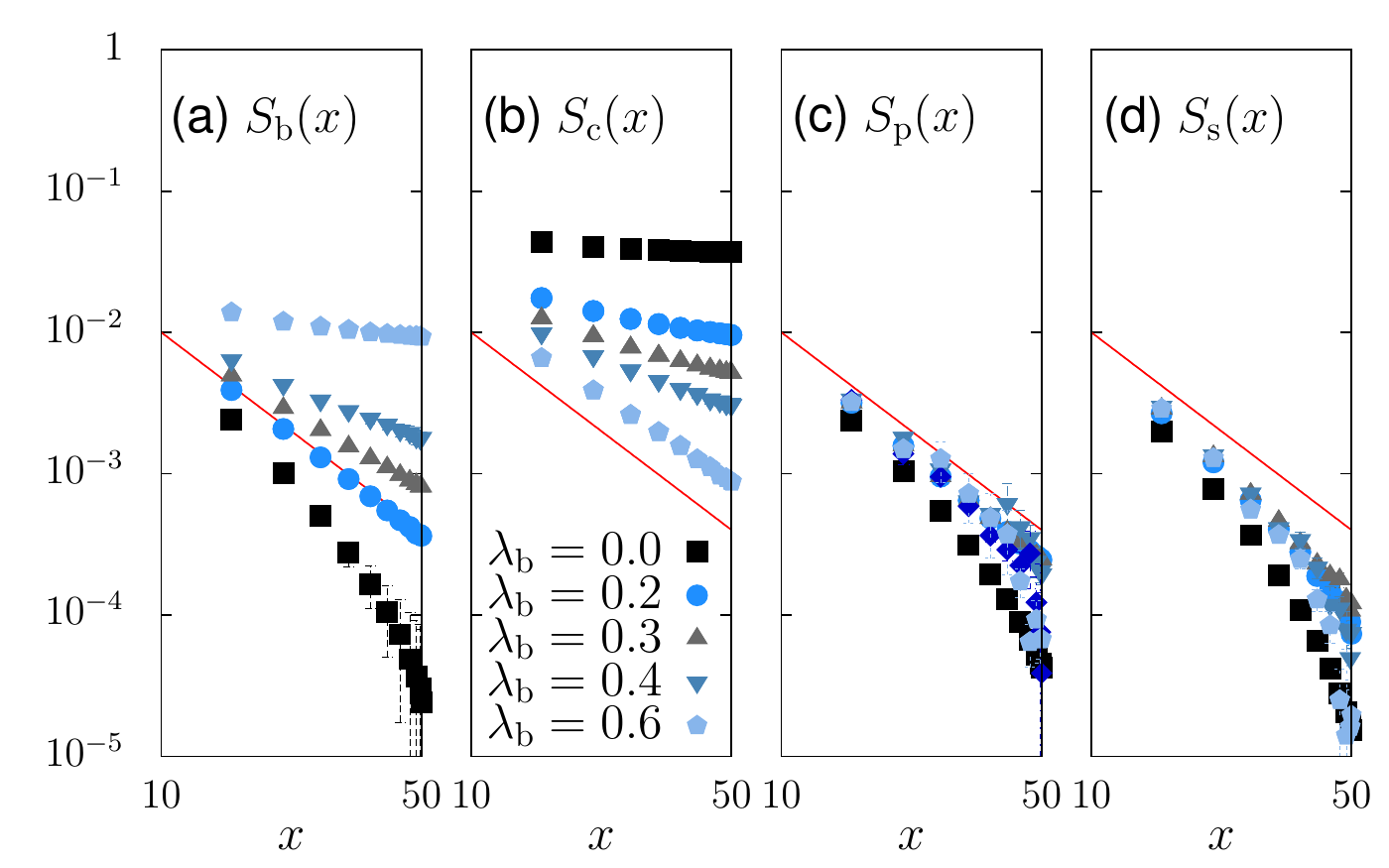}
\caption{\label{fig:sshhol-correlators}
(a) Bond, (b) charge, (c) pairing, and (d) spin correlations of the Holstein-SSH model~(\ref{eq:sshhol})
 from  QMC simulations \cite{PhysRevLett.117.206404}. Here, $\omega_0/t=0.5$. The solid line
indicates $1/x^2$.}
\end{figure}

Of particular interest is the question if the metallic phase of the Holstein
model is stable with respect to the SSH coupling, or if metallic behavior
is entirely absent as in the SSH model~(\ref{eq:ssh}). In terms of g-ology, 
both couplings produce negative backscattering matrix elements that give rise
to a spin gap. On the other hand, the umklapp matrix elements have opposite
sign and can therefore compensate, allowing for an extended LEL (C1S0) metallic
region. This picture is confirmed by QMC data \cite{PhysRevLett.117.206404} summarized in the qualitative phase diagram in
Fig.~\ref{fig:sshhol}. If the SSH coupling dominates, the system is a BOW
insulator just like the SSH model. If the Holstein coupling
dominates, a CDW ground state exists. Both states are of type C0S0.
If the couplings are comparable, the competition between the two orders
results in a metallic LEL phase. Starting in the CDW phase and increasing
$\lambda_\text{b}$, the correlation functions in
Fig.~\ref{fig:sshhol-correlators} reveal a suppression (enhancement) of CDW
(BOW) order and a QPT to the LEL phase with power-law correlations. At stronger SSH couplings,
long-range BOW order emerges. For all parameters, spin correlations remain
exponential due to the spin gap. The QPT between the two different
Peierls states is found to be continuous, and in the adiabatic regime
involves two soft-mode transitions for the site and bond phonon modes,
respectively \cite{PhysRevLett.117.206404}. The single-particle gap is
minimal but finite at the QPT \cite{PhysRevLett.117.206404}. These
numerical results contradict earlier approximate results suggesting a
first-order BOW--CDW transition \cite{PhysRevB.28.4833} or a ferroelectric
phase with coexistence of BOW and CDW order \cite{0953-8984-2-16-003}.

\subsection{Edwards model}\label{sec:edwards}

The discussion so far has revealed that the coupling to the lattice can modify the
transport properties of low-dimensional systems to the point of insulating
behavior. Quantum transport in general takes place in some ``background'',
which may consist of lattice but also spin or orbital degrees of freedom. For instance, a key problem in the widely studied 
high-$T_c$ cuprates~\cite{BM86} and colossal magnetoresistance
manganites~\cite{JS50} is that of (doped) holes moving in an ordered
magnetic insulator~\cite{Be09}. As the holes move, they disrupt the 
order of the background which, conversely, hinders hole motion. Coherent motion may still occur, albeit on a reduced
energy scale determined by the fluctuations and correlations in the background. 

A fermion-boson model by Edwards describes the interaction of particles with
the background in terms of a coupling to bosonic degrees of
freedom~\cite{Ed06,AEF07}:
\begin{align}\label{eq:edwardsmodel}
 \hat{H}
  = 
  -t\sum_{\langle i, j \rangle} \hat{c}_j^{\dagger} \hat{c}_{i}^{}
    (\hat{b}_i^{\dagger}+\hat{b}_j^{}) +\omega_0\sum_i \hat{b}_i^{\dagger}\hat{b}_i^{}
  -\bar{\lambda}\sum_i(\hat{b}_i^{\dagger}+\hat{b}_i^{})\,.
\end{align}
In this model, every hop of a (spinless fermionic) charge carrier along a 1D
transport path either creates a (local bosonic) excitation with energy
$\omega_0$ in the background at the site it leaves, or annihilates 
an existing excitation at the site it enters. The fermion-boson coupling in
Eq.~(\ref{eq:edwardsmodel}) differs significantly from the Holstein and SSH
couplings discussed before. In particular, no static distortion arises in the
limit $\omega_0\to 0$. Furthermore, spontaneous boson creation and
annihilation processes are possible, i.e., the background distortions can
relax with a relaxation rate $\bar{\lambda}$, for example due to quantum
fluctuations. Any particle motion is affected by the background and vice
versa. In fact, the Edwards model describes three different regimes:
quasi-free, diffusive, and boson-assisted transport~\cite{AEF07}. In the
latter case, excitations of the background are energetically costly
($\omega_0/t >1$) and the background relaxation rate is small
($\bar{\lambda}/t \ll 1$), i.e., the background is ``stiff''. Then, for a
half-filled band, strong correlations can develop and even drive the system 
into an insulating state by establishing long-range CDW order~\cite{WFAE08,EHF09,EF09b}.    

\begin{figure}
\centering
\includegraphics[width=0.45\textwidth]{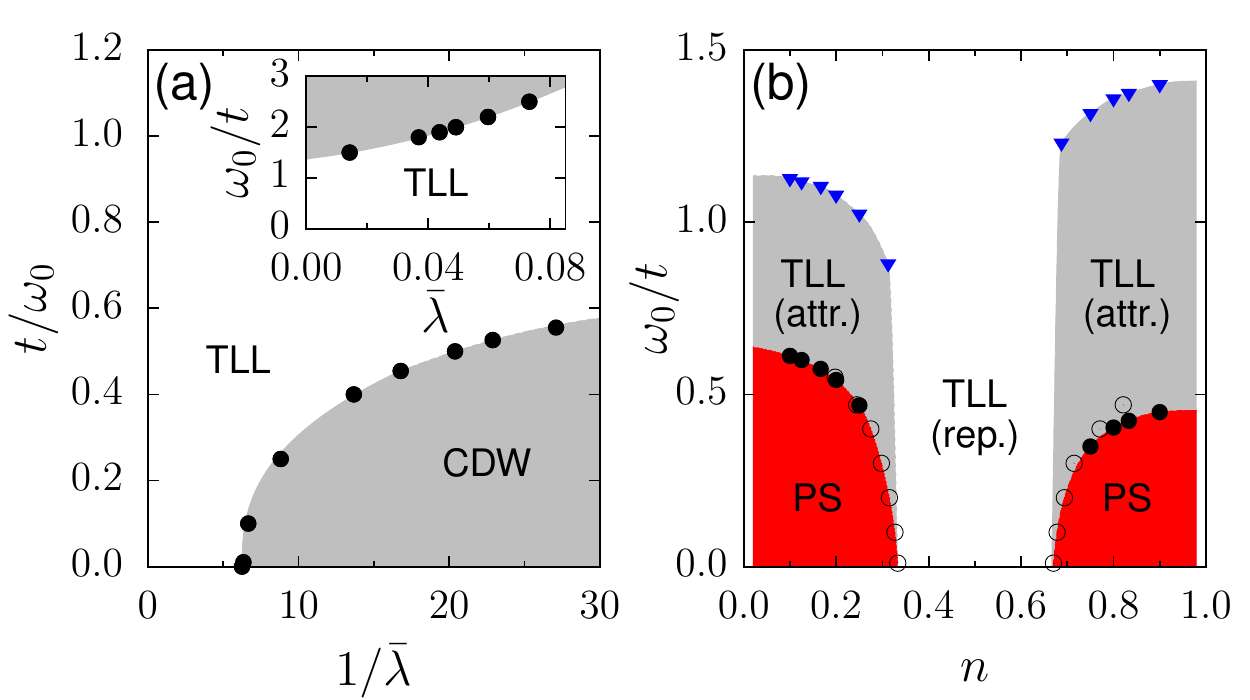}
\caption{\label{fig:edwards_phases}
(a) DMRG phase diagram of the Edwards model~(\ref{eq:edwardsmodel}) at half-filling~\cite{EHF09}. (b) Ground states as a
function of band filling and $\omega_0/t$ in the slow-boson regime at fixed
$\bar{\lambda}/t=0.2$, including regions of phase separation (PS)~\cite{ESBF12}.
}
\end{figure}

Figure~\ref{fig:edwards_phases}(a) shows the DMRG phase diagram at
half-filling. The critical values were determined from the charge gap and
the CDW order parameter~\cite{EHF09}. The metallic phase found below a critical boson
frequency and above a critical relaxation rate is a repulsive TLL
($K_\text{c}<1$)~\cite{EHF09}. Remarkably, particle motion is possible even for
$\bar{\lambda}=0$, in lowest order by a vacuum-restoring six-step process
where 3 bosons are excited in steps 1-3 and afterwards consumed in steps 4-6
with the particle moving two sites~\cite{AEF07}. In contrast to the spinless Holstein model, the
CDW state of the Edwards model is a few-boson
state~\cite{WFAE08}. As shown in Fig.~\ref{fig:edwards_phases}(b), at low and
high band filling $n$ the attractive interaction mediated by the slow bosons
becomes strong enough to give rise to first an attractive TLL ($K_\text{c}>1$)
and finally electronic phase separation~\cite{ESBF12}.
 
\begin{figure}[t]
\centering
\includegraphics[width=0.45\textwidth]{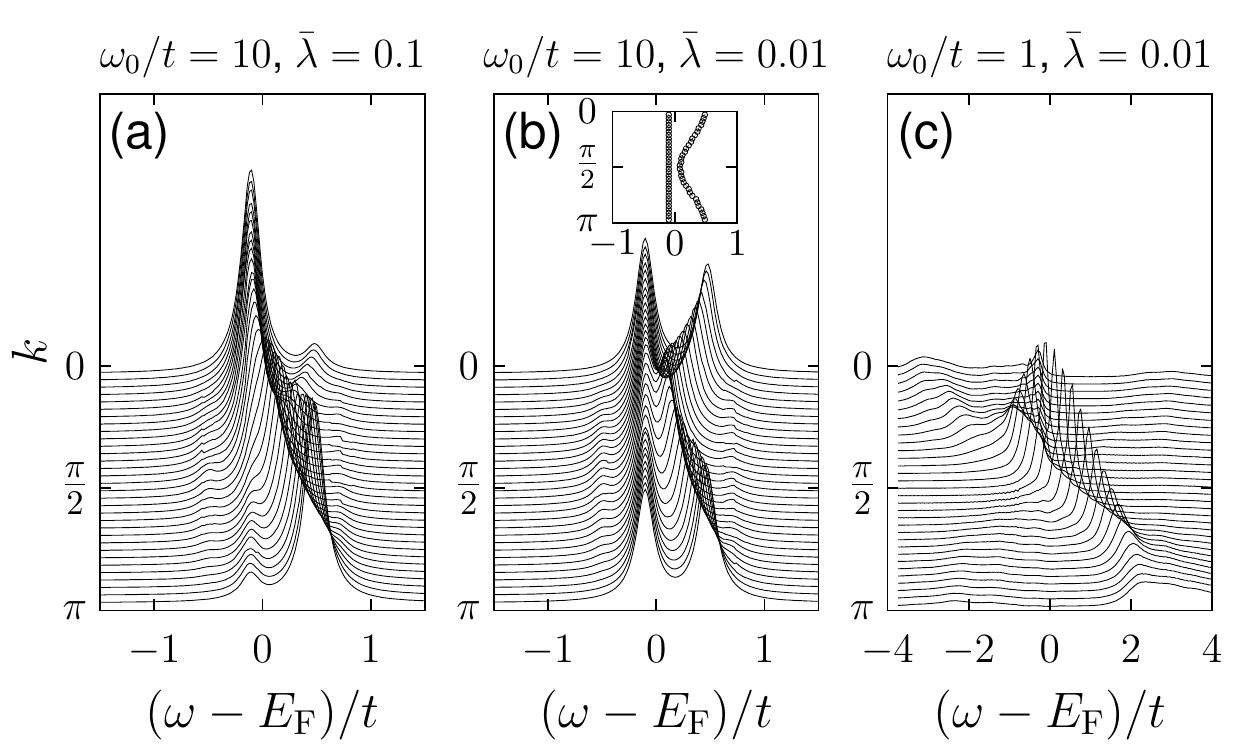}
\caption{\label{fig:edwards_spectra}
Single-particle spectral function of the Edwards
model~(\ref{eq:edwardsmodel}) at half-filling
from the dynamical DMRG~\cite{FEWB12}. Inset: dispersion of
the absorption/emission maximum.}
\end{figure}

The metal--insulator QPT at half-filling is also reflected in the
photoemission spectra measured by 
\begin{equation}\label{spsfpm}
  A^{\pm}(k,\omega)
  =\sum_n|\langle\psi_n^{\pm}|\hat{c}^{\pm}_k|\psi_0\rangle|^2\,
  \delta[\omega\mp(E_n^\pm-E_0)]\,,
\end{equation}
where $\hat{c}^+_k=\hat{c}^\dagger_k$, $\hat{c}^-_k=\hat{c}^{}_k$, $|\psi_0\rangle$ 
is the ground state for $N_\text{e}$ particles and 
$| \psi_n^{\pm}\rangle$ the $n$-th  excited state with $N_\text{e}\pm 1$ particles.
Figure~\ref{fig:edwards_spectra} shows $A(k,\omega)=A^-(k,\omega)+A^+(k,\omega)$
in the regime where
background excitations have a large energy and the bosons strongly affect
particle transport. The  quasiparticle mass is significantly enhanced and a renormalized band structure 
appears. However, if $\bar{\lambda}$ is sufficiently large, the system remains metallic,
as indicated by a finite spectral weight at the Fermi energy $\EF$
[Fig.~\ref{fig:edwards_spectra}(a)]. As the possibility of relaxation reduces
a gap opens at $\kF=\pi/2$ [Fig.~\ref{fig:edwards_spectra}(b)], indicating
insulating behavior. We note the internal feedback 
mechanism: The collective boson excitations originate from the motion of the 
charge carriers and have to persist long enough to finally inhibit particle transport,
thereby completely changing the nature of the many-particle ground state.
This complex boson-particle dynamics leads to a new, correlation-in{\-}duced band
structure reminiscent of the (extended) Fali{\-}cov-Kim{\-}ball model, with a very
narrow valence band and a rather broad conduction band [inset of Fig.~\ref{fig:edwards_spectra}(b)]. The
asymmetric masses can be understood by ``doping'' a
perfect CDW state:  To restore the CDW order a doped hole has to be
transferred by the above-mentioned six-step process of order ${\cal
  O}(t^6/\omega_0^5)$, while a doped particle can move by a two-step process of order ${\cal O}(t^2/\omega_0)$~\cite{WFAE08}.
By decreasing $\omega_0$ at fixed $\bar{\lambda}$ the fluctuations overcome the correlations and the system returns
to a metallic state. However, the latter differs from the state we
started with. In particular, $A(k,\omega)$ in Fig.~\ref{fig:edwards_spectra}(c)
shows sharp absorption features only near $\kF$ and ``overdamping'' at the zone boundaries where 
the spectrum is dominated by bosonic excitations. 

\subsection{Heisenberg spin-Peierls models}\label{sec:spinpei}

The Peierls (dimerization) instability triggered by the lattice degrees of
freedom can be observed not only in quasi-1D itinerant electron systems but
also in spin chains with magneto-elastic couplings. Experimentally, such
behavior was first seen in the 1970s for organic compounds of the TTF and
TCNQ family~\cite{Brea75}. Interest in the subject revived after the
discovery of the first inorganic spin-Peierls compound $\rm CuGeO_3$ in
1993~\cite{HTU93}, in particular due to the fact that the displacive
spin-Peierls transition in this material does not involve phonon softening.
Instead, the Peierls-active optical phonon modes with frequencies
$\omega_{0,1}\simeq J$ and $\omega_{0,2}\simeq 2J$ ($J$ being the exchange
coupling between neighboring $\rm Cu^{2+}$ ions that  form well separated
spin-$\oh$ chains) harden by about 5\% at the transition which therefore 
occurs at very strong spin-phonon coupling~\cite{BHRDR98}.  Phonon
hardening for experimentally relevant parameters was demonstrated for the
magnetorestrictive XY model by calculating the dynamic structure
factor~\cite{HFW01}. The physics of $\rm CuGeO_3$ reveals that the canonical
adiabatic treatment of the lattice~\cite{Py74,CF79} is inadequate for this material~\cite{BFKW99}.
Instead, the application of numerical methods to paradigmatic quantum models
yields key information about the nature of the phase transition and the
correct models for inorganic spin-Peierls materials.

The simplest model containing all important features of a spin-Peierls
system is an antiferromagnetic Heisenberg chain,
$\hat{H}_{\rm Heis}= J \sum_i \hat{{\bm S}}_i \cdot \hat{{\bm S}}_{i+1}$ ($J<0$, $\hat{{\bm S}}_i$ is a
spin-$\oh$ operator at site $i$), coupled to Einstein quantum phonons:
\begin{align}\label{eq:spinpei}
\hat{H}=\hat{H}_{\rm Heis} + \hat{H}_{\rm SP}^{(\rm l,d)}+ \omega_0\sum_i
  \hat{b}^{\dagger}_i \hat{b}^{\phantom{\dagger}}_{i}\,.
\end{align}
Here we consider two different spin-phonon couplings,
\begin{align}\label{eq:li}
\hat{H}_{\rm SP}^{\rm l}&= g\omega_0\sum_i (\hat{b}_i^\dagger + \hat{b}_i^{\phantom{\dagger}})\, \hat{{\bm S}}_i\cdot \hat{{\bm S}}_{i+1}\,,\\\label{eq:di}
\hat{H}_{\rm SP}^{\rm d}&= g\omega_0\sum_i (\hat{b}_{i+1}^{\dagger}+\hat{b}_{i+1}^{\phantom{\dagger}} - \hat{b}_{i}^{\dagger} -\hat{b}_{i}^{\phantom{\dagger}}  )\, \hat{{\bm S}}_i\cdot \hat{{\bm S}}_{i+1}\,.
\end{align} 
The local coupling $\hat{H}_{\rm SP}^{\rm l}$ captures the modification
of the spin exchange by a local lattice degree of freedom (modeling, \eg,
side-group effects)~\cite{WFK98,WWF99}. The difference coupling $\hat{H}_{\rm
 SP}^{\rm d}$ describes a linear dependence of the spin exchange on the
difference between the phonon amplitudes at sites $i$ and $i+1$~\cite{BMH99}.   
A first insight into these models can be gained by integrating out the
phonons in the anti-adiabatic limit $\omega_0\gg J$ to obtain an
effective Heisenberg model with longer-ranged interactions that give rise
to frustration. The spin Hamiltonian  $\hat{H}= J
\sum_i (\hat{{\bm S}}_i \cdot \hat{{\bm S}}_{i+1} + \alpha \hat{{\bm S}}_i
\cdot \hat{{\bm S}}_{i+2})$ has a dimerized ground state (alternating long
and short bonds) for $\alpha \geq
\alpha_c=0.241\,167$~\cite{ON92}. Accordingly, the spin-phonon
coupling must be larger than a nonzero critical value $g_c(\omega_0)$ for
the spin-Peierls instability to occur~\cite{Uh98,WWF99,BMH99}. This is
similar to the Holstein model~\eqref{eq:holstein} but in contrast to the
static limit $\omega_0/J=0$. 

\begin{figure}
\centering
\includegraphics[width=0.45\textwidth]{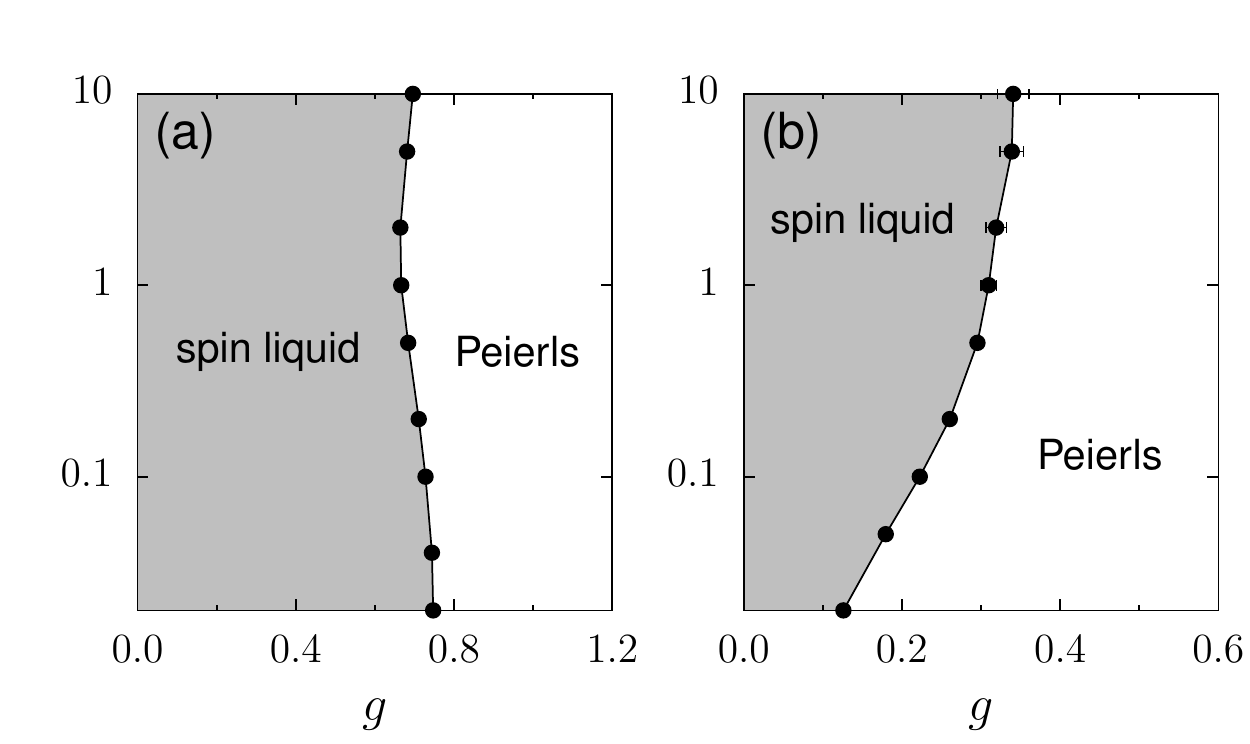}
\caption{\label{fig:spinpei-pd}
DMRG phase diagrams of the Heisenberg spin-Peierls model~(\ref{eq:spinpei})
with (a) a local coupling $\hat{H}_{\rm SP}^{\rm l}$ \cite{WHBF06} and (b) a difference
coupling $\hat{H}_{\rm SP}^{\rm d}$ \cite{BMH99}.}
\end{figure}

Figure~\ref{fig:spinpei-pd} shows the phase diagram of the
model~\eqref{eq:spinpei} for either the coupling~(\ref{eq:li}) or~(\ref{eq:di}) from
two-block \cite{BMH99} and four-block~\cite{WHBF06} DMRG calculations,
respectively.  The QPT
from the the gapless spin-liquid state to the gapped dimerized phase was
detected using the well-established criterion of a level crossing between the
first singlet and the first triplet excitation; the latter was derived for
the frustrated spin chain~\cite{AGSZ89,BMH99,HWWJF07}. For finite systems,
the singlet lies above the triplet excitation in the spin liquid,
and the two levels become degenerate with the singlet ground state as $L\to
\infty$. In the symmetry-broken gapped phase, the lowest singlet state becomes degenerate
with the ground state. The Heisenberg spin-Peierls model with quantum phonons
is in the same universality class as the frustrated spin chain
\cite{BMH99}. The phonon spectral function has been analyzed in \cite{arXiv:0705.0799}. Related spin-boson models exhibiting TLL and CDW phases 
have also been investigated in the context of dissipative quantum systems
\cite{PhysRevLett.113.260403}.

\section{Density waves and Coulomb interaction}\label{sec:coulomb}

\subsection{Holstein-Hubbard model}\label{sec:holsteinhubbard}

From the 1D Hubbard model \cite{Hu63}, it is well established that a local Coulomb repulsion favors
a correlated MI with dominant $2\kF$ SDW fluctuations \cite{Giamarchi}. In
contrast to CDW and BOW order, the continuous SU(2) spin symmetry cannot be
spontaneously broken \cite{PhysRevLett.17.1133}. Instead, the SDW correlations are critical
($\sim 1/r$) \cite{Schulz90,Giamarchi}. Of key interest is the interplay or
competition of retarded electron-phonon and instantaneous electron-electron
interactions that determines if the ground state is a CDW/BOW, SDW, or LEL state.
A minimal but rich model capturing this interplay is the Holstein-Hubbard 
model with Hamiltonian 
\begin{align}\label{eq:hhm}
 \hat{H} =       \hat{H}_{\rm Hol}  + U
      \sum\limits_i \on_{i\uparrow} \on_{i\downarrow}\,.
\end{align}

The ground-state phase diagram of Eq.~(\ref{eq:hhm}) was the subject of intense debate. 
Even after early claims \cite{Hirsch83a} (unfounded \cite{PhysRevB.92.245132}
but supported by RG calculations \cite{PhysRevB.71.205113,Ba.Bo.07}) of the absence of metallic behavior in the spinful
Holstein model were contradicted by DMRG
results \cite{JeZhWh99}, numerical work on the Holstein-Hubbard model
initially focused on strong couplings where a direct SDW--CDW QPT is observed.
Evidence for an intermediate metallic phase---expected from the adiabatic connection
to the Holstein model as $U\to0$---at weaker couplings was obtained with a 
variety of different methods~\cite{TC03,TAA05,TAA07,CH05,HC07,FHJ08,EF10}. Very recently, it was
shown that the absence of such a phase in RG calculations \cite{PhysRevB.71.205113,Ba.Bo.07} is due to
the neglected momentum dependence of the interaction \cite{Barkim2015}. A detailed
discussion of these contradictory findings was given in \cite{PhysRevB.92.245132}.

\begin{figure}[t]
\centering
\includegraphics[width=0.45\textwidth]{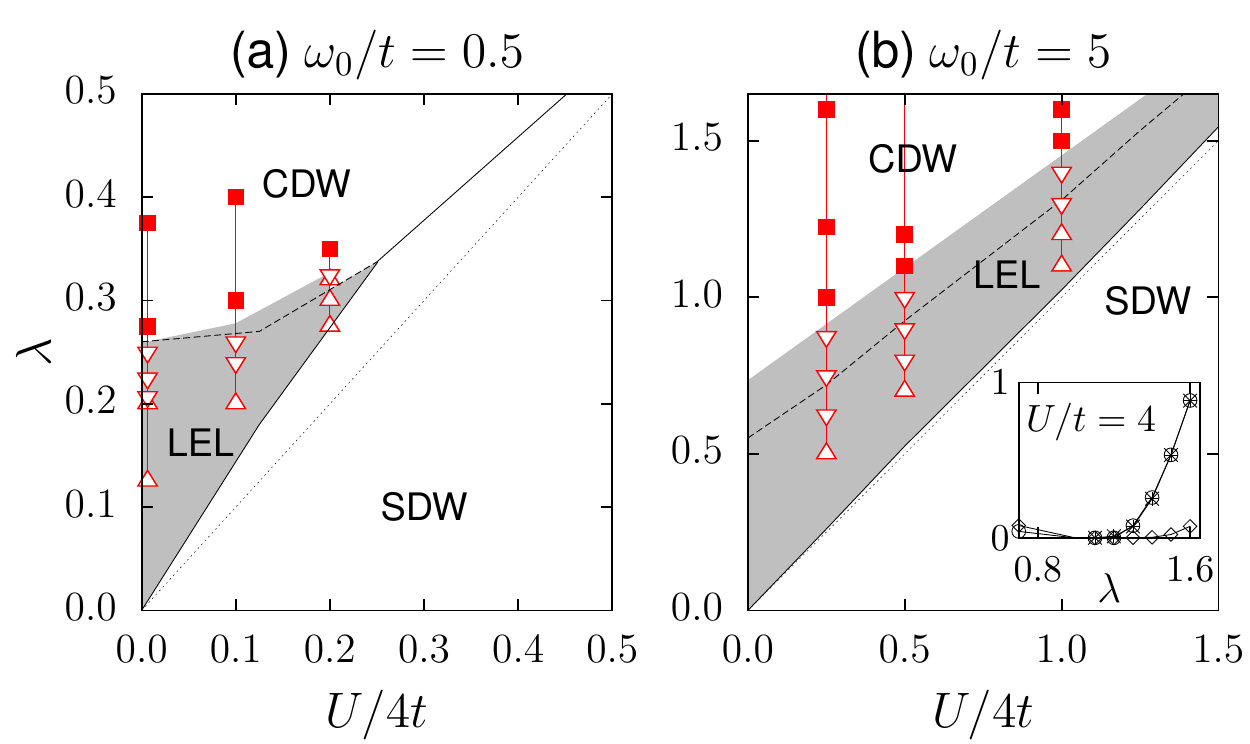}
\caption{\label{fig:holhub_phasediagram}
DMRG phase diagram of the Holstein-Hubbard model~(\ref{eq:hhm}) in (a)
the adiabatic and (b) the anti-adiabatic regime~\cite{FHJ08}. Dashed (solid)
lines are CDW--LEL (LEL--SDW) critical values from QMC~\cite{CH05}, the
dotted line is $U=4\lambda t$. Squares (triangles) indicate the CDW Peierls
(LEL) phase. The inset shows the one-particle (circles), two-particle
(diamonds), and spin (stars) excitation gaps in the thermodynamic limit.}
\end{figure}

\begin{figure}
\centering
\includegraphics[width=0.45\textwidth]{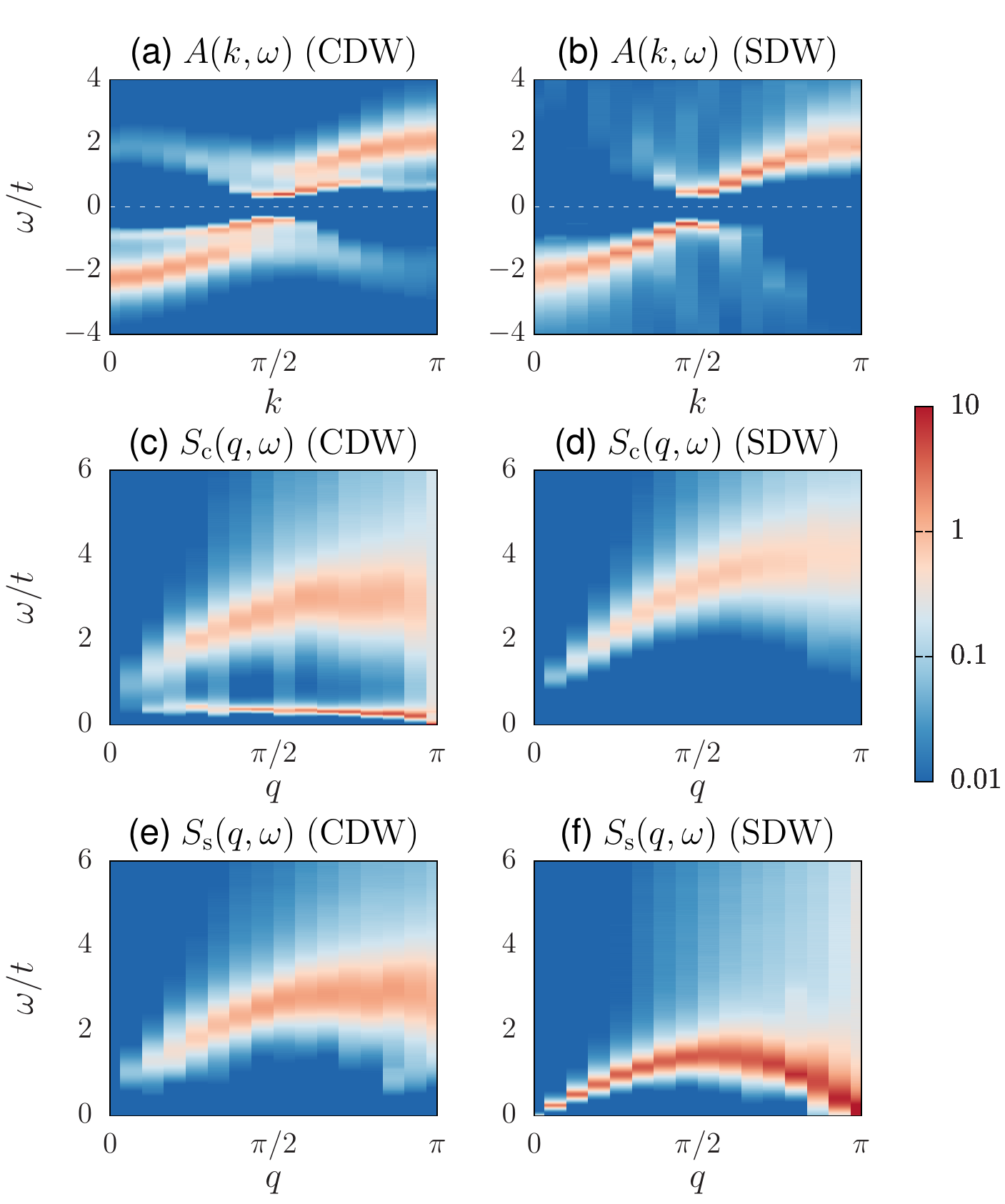}
\caption{\label{fig:holhub_spectra}
Single-particle [(a),(b)], density [(c),(d)], and spin [(e),(f)] excitation
spectra of Holstein-Hubbard model~(\ref{eq:hhm}) from QMC
\cite{PhysRevB87.075149} in the CDW ($\omega_0/t=0.5$, $U/t=0.2$,
$\lambda=0.4$) and the SDW phase ($\omega_0/t=5$, $U/t=4$, $\lambda=0.25$).
}
\end{figure}

The currently most reliable phase boundaries come from DMRG \cite{FHJ08,EF10}
and QMC \cite{CH05,HC07} calculations. The DMRG critical values shown in
Fig.~\ref{fig:holhub_phasediagram} were obtained from a finite-size scaling of the single-particle, charge-, spin-,
and neutral gaps defined as 
\begin{align}\label{eq:gaps} \nonumber
\Delta_{\text{c}_1}&=E_0(L+1,\mbox{$\frac{1}{2}$})+E_0(L-1,\mbox{$\frac{1}{2}$})-2E_0(L,0)\,,\\\nonumber
\Delta_{\text{c}_2}&=E_0(L+2,\,0)+E_0(L-2,0)-2E_0(L,\,0)\,,\\\nonumber
\Delta_{\text{s}\,\,} &= E_0(L,1)-E_0(L,0)\,,\\
\Delta_{\text{n}\,\,} &= E_1(L,0)-E_0(L,0)\,.
\end{align}
$E_0(N_\text{e},S^z_\text{tot})$ [$E_1(N_\text{e},S^z_\text{tot})$] is the energy of the
ground-state (first excited state) of a system with $L$ sites, $N_\text{e}$
electrons and total spin-$z$ $S_\text{tot}^z$.  The CDW state has C0S0
($\Delta_{\text{c}1}>0$, $\Delta_\text{s}>0$),  whereas the
SDW state has C0S1 ($\Delta_{\text{c}1}>0$, $\Delta_\text{s}=0$). The
different nature of excitations in these phases is clearly visible
in the spectra in Fig.~\ref{fig:holhub_spectra} (for
previous work see \cite{Voit98,FeWeHaWeBi03,WeAsHo15I}). The single-particle
spectral function in Figs.~\ref{fig:holhub_spectra}(a)
and (b) has a gap at $\EF$ in both phases, but distinct soliton excitations and
backfolded shadow bands only in the CDW phase. Spin-charge separation
\cite{Giamarchi} can be observed for strong interactions
\cite{PhysRevB.75.205128,NiZhWuLi05}. The dynamic charge structure factor
\begin{align}\label{eq:nqw}
  S_\text{c}(q,\om)
  &=
  \frac{1}{Z}\sum_{mn} {|\bra{\psi_m} \hat{\rho}_q \ket{\psi_n}|}^2
  e^{-\beta E_m} 
  \delta[\om - (E_n - E_m)]
\end{align}
with $\hat{\rho}_q =  \sum_r e^{iqr} (\on_{r} - n)/\sqrt{L}$ 
in Figs.~\ref{fig:holhub_spectra}(c),(d) reveals a $q=0$ charge gap 
in both insulating phases and the renormalized phonon frequency in the CDW phase
[Fig.~\ref{fig:holhub_spectra}(c)]. Finally, the dynamic spin structure factor
\begin{equation}\label{eq:sqw}
  S_\text{s}(q,\om)
  =
  \frac{1}{Z}\sum_{mn} {|\bra{\psi_m} \hat{S}^z_q \ket{\psi_n}|}^2
  e^{-\beta E_m} 
  \delta[\om - (E_n - E_m)]
\end{equation}
shows a clear spin gap in the CDW phase [Fig.~\ref{fig:holhub_spectra}(e)] whereas
the SDW phase has $\Delta_\text{s}=0$ and strong $2\kF=\pi$ fluctuations [Fig.~\ref{fig:holhub_spectra}(f)].

Similar to the Holstein model, the intermediate phase is a spin-gapped LEL
(C1S0, $\Delta_{\text{c}1}>0$, $\Delta_\text{s}>0$, but $\Delta_{\text{c}2}=0$)
\cite{HC07,FHJ08,PhysRevB87.075149,Barkim2015}. 
In the anti-adiabatic regime [Fig.~\ref{fig:holhub_phasediagram}(b)],
where retardation effects are small, the LEL--SDW QPT occurs close to
the value $U=4\lambda t$ expected from an effective Hubbard
model. 
Whereas DMRG and QMC results agree quite well for the LEL--SDW QPT,
the LEL--CDW QPT line is not completely settled. The exponential opening of the charge gap is nontrivial to detect
with the DMRG, and the charge susceptibility used in QMC~\cite{CH05} is problematic due
to the spin gap \cite{PhysRevB.92.245132}. The latter also complicates the calculation of
TLL parameters \cite{PhysRevB.92.245132}. The intermediate LEL phase has $K_\text{s}=0$, so
that the low-energy theory is that of bosonic pairs (bipolarons). $K_\text{c}$ as extracted
from the electronic density structure factor gives $K_\text{c}>1$ even though
pairing correlations are always subdominant \cite{Barkim2015} (see also
Fig.~\ref{fig:holhub_correlators}). Moreover, in contrast to the extended
Hubbard model, there is no symmetry argument for $K_\text{c}=1$ at the LEL--CDW QPT.
An interesting open problem is to reconcile the vanishing of the
(bipolaron) binding energy in parts of the LEL phase \cite{EF10} with the
nonzero spin gap. Finally, CDW and SDW states of the Holstein-Hubbard model
have been studied numerically in the context of pump-probe experiments
\cite{JPSJ.81.013701,PhysRevB.79.125118,PhysRevB.84.195109,PhysRevB.88.064303,PhysRevLett.109.246404,PhysRevLett.109.176402,PhysRevLett.116.086401}.

\subsection{SSH-$UV$ model}\label{sec:sshuv}

\begin{figure}[t]
\centering
\includegraphics[width=0.45\textwidth]{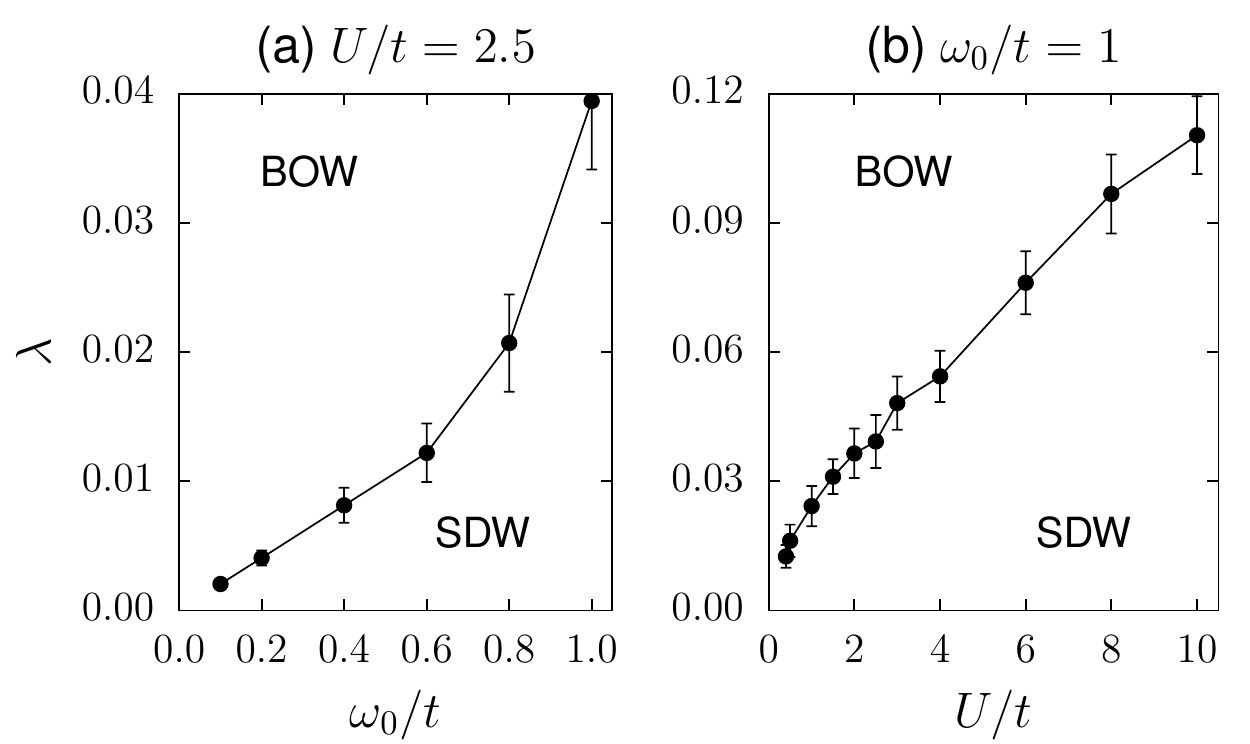}
\caption{\label{fig:sshuv}
Phase diagrams of the SSH-$UV$ model~(\ref{eq:sshuv}) from QMC
simulations for $V=U/4$. Data taken from \cite{PhysRevB.67.245103}.}
\end{figure}

The competition between electron-phonon and electron-electron interaction has also been studied in the
framework of the SSH-$UV$ Hamiltonian
\begin{equation}\label{eq:sshuv}
  \hat{H} = \hat{H}_\text{SSH} + U \sum\limits_i \on_{i\uparrow} \on_{i\downarrow}
  + V \sum\limits_i \on_{i} \on_{i+1}\,,
\end{equation}
which is directly relevant for conjugated polymers \cite{Baeriswylreview92,PhysRevB.57.11838}.
The phase diagram from QMC simulations \cite{PhysRevB.67.245103} is shown in Fig.~\ref{fig:sshuv}.
A key difference to the Holstein-Hubbard model is that no metallic phase results from
the competing interactions. Instead, for $\omega_0>0$ and realistic values
$U>2V$, the ground state is a MI with critical SDW correlations (C0S1)
for $\lambda<\lambda_c$, and a BOW Peierls state (C0S0) for
$\lambda>\lambda_c$~\cite{CB84,PhysRevB.67.245103,PhysRevB.83.195105,PhysRevB.91.245147};
in contrast to the continuous suppression of CDW correlations
by the Hubbard repulsion in the Holstein-Hubbard model, the amplitude of BOW
correlations is enhanced by the Coulomb repulsion in the SSH-$UV$ model
\cite{PhysRevB.26.4278,PhysRevLett.51.292,PhysRevLett.51.296,PhysRevLett.62.1053,PhysRevB.91.245147}.
Finally, for large $U$, the SSH-$UV$ model is closely related to the spin-Peierls models
discussed in Sec.~\ref{sec:spinpei} \cite{PhysRevB.67.245103}.

\subsection{Extended Falicov-Kimball model}\label{sec:efkm}

CDW, BOW, SDW or orbital DW states can also arise purely from the Coulomb
interaction, so that extensions of the Hubbard Hamiltonian may be
regarded as minimal theoretical models.  An important example is the asymmetric Hubbard
model with spin-dependent band energies $\varepsilon_{k\sigma}=E_\sigma - 2 t_\sigma \cos k$, where $E_\sigma$ defines the center 
of the spin-$\sigma$ band and $t_\sigma$ is the nearest-neighbor hopping
amplitude~\cite{LD94,Fa08}. For $E_\uparrow < E_\downarrow$ and  $t_\uparrow t_\downarrow<0$ ($t_\uparrow t_\downarrow>0$) a direct 
(indirect) band gap is realized. The $\sigma$-electron density
$n_\sigma  =\frac{1}{L} \sum_{k} \langle \hat{c}_{k\sigma}^\dagger \hat{c}_{k\sigma}^{} \rangle$, with $n_\uparrow+n_\downarrow=1$
at half-filling. The asymmetric Hubbard model has been used to investigate various
many-body effects in (mixed/intermediate-valence) rare-earth and
transition-metal compounds, including the DW--PS QPT~\cite{WCG75}, electronic ferroelectricity~\cite{Ba02b,YMDLH03} and  
(pressure-induced) exciton condensation \cite{IPBBF08}, as well as
multiorbital correlation physics in cold atoms~\cite{WCR12}. Regarding
$\sigma$ as an orbital flavor, the asymmetric Hubbard
model is equivalent to the extended Falicov-Kimball model
(EFKM)~\cite{LD94,Fa08,Ba02b,ZIBF12} 
\begin{align}\label{eq:hamilEFKM}
 \hat{H}_{\rm EFKM}= &-t_c\sum_{\langle i, j \rangle}
               \hat{c}_{i}^{\dagger} \hat{c}_{j}^{\phantom{\dagger}}
         -t_f\sum_{\langle i, j \rangle}
               \hat{f}_{i}^{\dagger} \hat{f}_{j}^{\phantom{\dagger}}             
\\\nonumber
         &+U\sum_i
                \hat{c}_{i}^{\dagger} \hat{c}_{i}^{\phantom{\dagger}}
                \hat{f}_{i}^{\dagger} \hat{f}_{i}^{\phantom{\dagger}}
+\frac{D}{2}\sum_i
                 \left(
                  \hat{c}_{i}^{\dagger} \hat{c}_{i}^{\phantom{\dagger}}
                 -\hat{f}_{i}^{\dagger} \hat{f}_{i}^{\phantom{\dagger}}
                 \right),
\end{align}
describing two species of spinless fermions, namely,  light $c$ (or $d$) electrons
and heavy $f$ electrons. A finite $f$-bandwidth allows for $f$-$c$ electron coherence, which will take account of a mixed-valence situation as well as 
of $c$-electron $f$-hole (exciton) bound-state formation and condensation~\cite{ZIBF12,Ku15}.
By contrast, $t_f=0$ in the original FKM~\cite{FK69}, so that the number of
$f$-electrons is strictly conserved  and no coherence between $f$ and $c$
electrons can arise \cite{SB88}.

\begin{figure}
\centering
\includegraphics[width=0.24\textwidth]{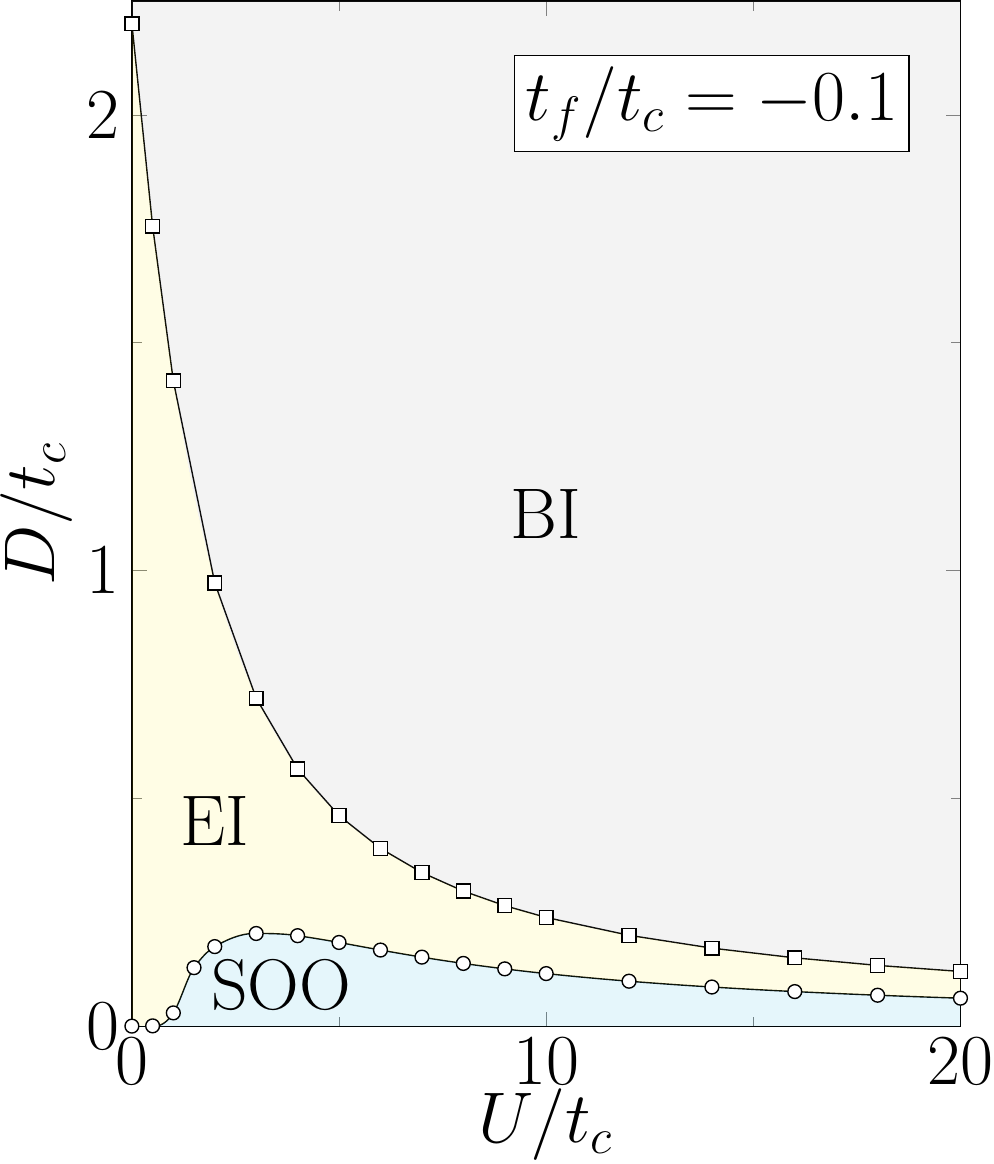}
\includegraphics[width=0.24\textwidth]{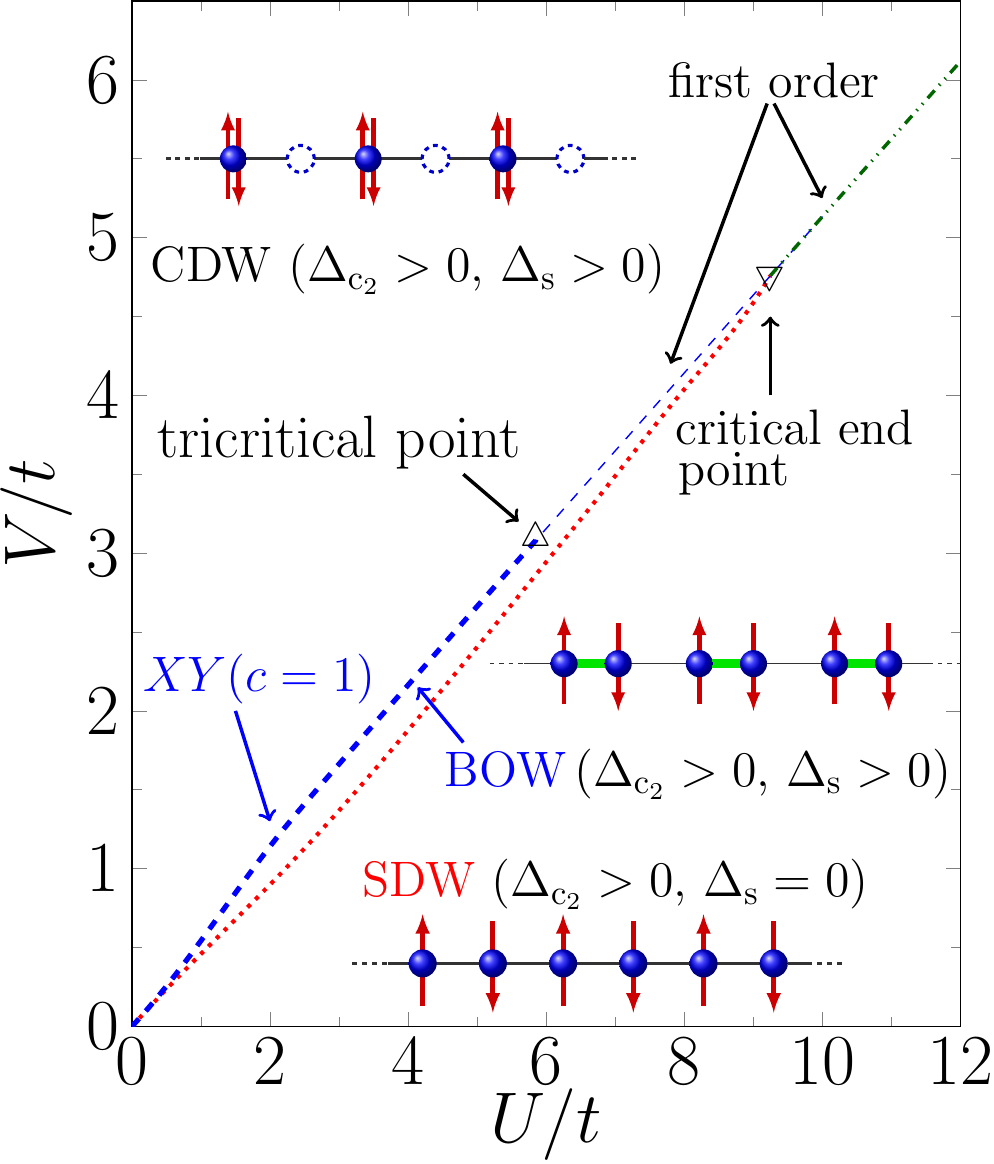}
\caption{\label{fig:efkm-ehm}
DMRG phase diagrams of the extended Falicov-Kimball model~(\ref{eq:hamilEFKM})
(left)~\cite{EKOF13} and the extended Hubbard model~(\ref{eq:ehm}) (right)~\cite{EN07}.}
\end{figure}

The left panel of Fig.~\ref{fig:efkm-ehm} shows the DMRG phase diagram at half-filling.
Depending on the orbital level splitting there exist staggered orbital
ordered (SOO) or band insulator (BI) phases, separated by a critical excitonic 
insulator (EI)~\cite{JRK67}. In the absence of true long-range order in one dimension,
the characteristic signatures of an excitonic Bose-Einstein condensate are a
power-law decay of the correlator $\langle \hat{X}_{i}^\dagger \hat{X}_{j}^{\phantom{\dagger}}\rangle$ with $\hat{X}_{i}^\dagger=\hat{c}_{i}^\dagger \hat{f}_{i}^{\phantom{\dagger}}$
and a divergence of the excitonic momentum distribution $N(q)=\langle \hat{X}_{q}^\dagger \hat{X}_{q}^{\phantom{\dagger}}\rangle$
with $\hat{X}_{q}^\dagger=\frac{1}{\sqrt{L}}\sum_k \hat{c}_{k+q}^\dagger \hat{f}_{k}^{\phantom{\dagger}}$ for the lowest-energy state (which has $q=0$ for
the case of a direct gap). The criticality of the EI can also be detected from the von Neumann entropy and the central charge [$c^*(L)\simeq 1$]. 
Monitoring the coherence length and binding energy with increasing $U$ yields
clear evidence for a BCS--BEC crossover \cite{EKOF13}. 
The addition of an electron-phonon coupling term to the EFKM leads to a competition between an ``excitonic'' CDW and
a ``phononic'' CDW, while an additional Hund's coupling promotes an excitonic SDW state~\cite{KZFO15}.

\subsection{Extended Hubbard model}\label{sec:extendedhubbard}

Another important and intensely investigated purely electronic model is the
extended Hubbard model (EHM)
\begin{align}\label{eq:ehm}
  \hat{H}_{\rm EHM}
  = 
  -t \sum_{\las i,j\ras\sigma} \hat{c}^\dagger_{i \sigma} \hat{c}_{j\sigma}^{\phantom{\dagger}}    
  +U \sum_i \on_{i \uparrow} \on_{i \downarrow} + V \sum_{i} \on_{i} \on_{i+1} \,.
\end{align}
It describes the competition between a local
Hubbard repulsion $U$ and a nonlocal (nearest-neighbor) repulsion $V$.
The phase diagram at half-filling has been determined by analytical~\cite{PhysRevB.45.4027,Na99,TF02,TTC06} and
numerical~\cite{Hi84,Je02,SBC04,PhysRevB.70.235107} methods. While there is agreement
that for $U\lesssim 2V$ ($U\gtrsim 2V$) the ground state has long-range
(critical) $2\kF$ CDW (SDW) correlations, the criticality of the QPTs
and the possibility of an intermediate BOW phase remain under
debate. The right panel of Fig.~\ref{fig:efkm-ehm} shows the currently
perhaps most accurate DMRG phase diagram~\cite{EN07}. The CDW phase is of
type C0S0, whereas the SDW phase has C0S1. Below a critical end point, they are
separated by a narrow C0S0 phase with long-range BOW
order~\cite{SBC04,TTC06,EN07}. Exactly on the CDW--BOW critical line
$\Delta_{\text{c}_2}= 0$ but $\Delta_{\text{c}_1}, \Delta_\text{s}>0$,
corresponding to an LEL (C1S0) \cite{PhysRevB.45.4027,Je02,SBC04,EN07}.
The CDW--BOW QPT changes from continuous (XY universality,
central charge $c=1$) to first order at the tricritical point $(U_\text{t}/t,V_\text{t}/t)
\simeq (5.89, 3.10)$~\cite{EN07}. The SDW--BOW QPT is characterized by
the opening of the spin gap. A detailed discussion of the low-energy theory and correlation functions
has been given in \cite{PhysRevB.45.4027}. Optical excitation spectra were calculated in \cite{PhysRevB.67.075106}.
 While it does not account for retardation effects, the EHM shares many of the features of the Holstein-Hubbard,
Holstein-SSH, and SSH-$UV$ models discussed in
Sec.~\ref{sec:fermion-boson}. Material-specific EHMs such
as H\"uckel-Hubbard-Ohno and Peierls-Hubbard-Ohno models have been studied in
detail with the DMRG method \cite{PhysRevB.87.245116,PhysRevB.95.085150}. Finally, a TLL to $4\kF$-CDW QPT as a
function of the Coulomb interaction range can be observed at quarter-filling \cite{PhysRevB.85.195115}.

\section{Density waves and symmetry protection}\label{sec:symmetryprotected}

\subsection{Dimerized Extended Hubbard model}\label{sec:ehmdimer}

\begin{figure}
\centering
\includegraphics[width=0.4\textwidth]{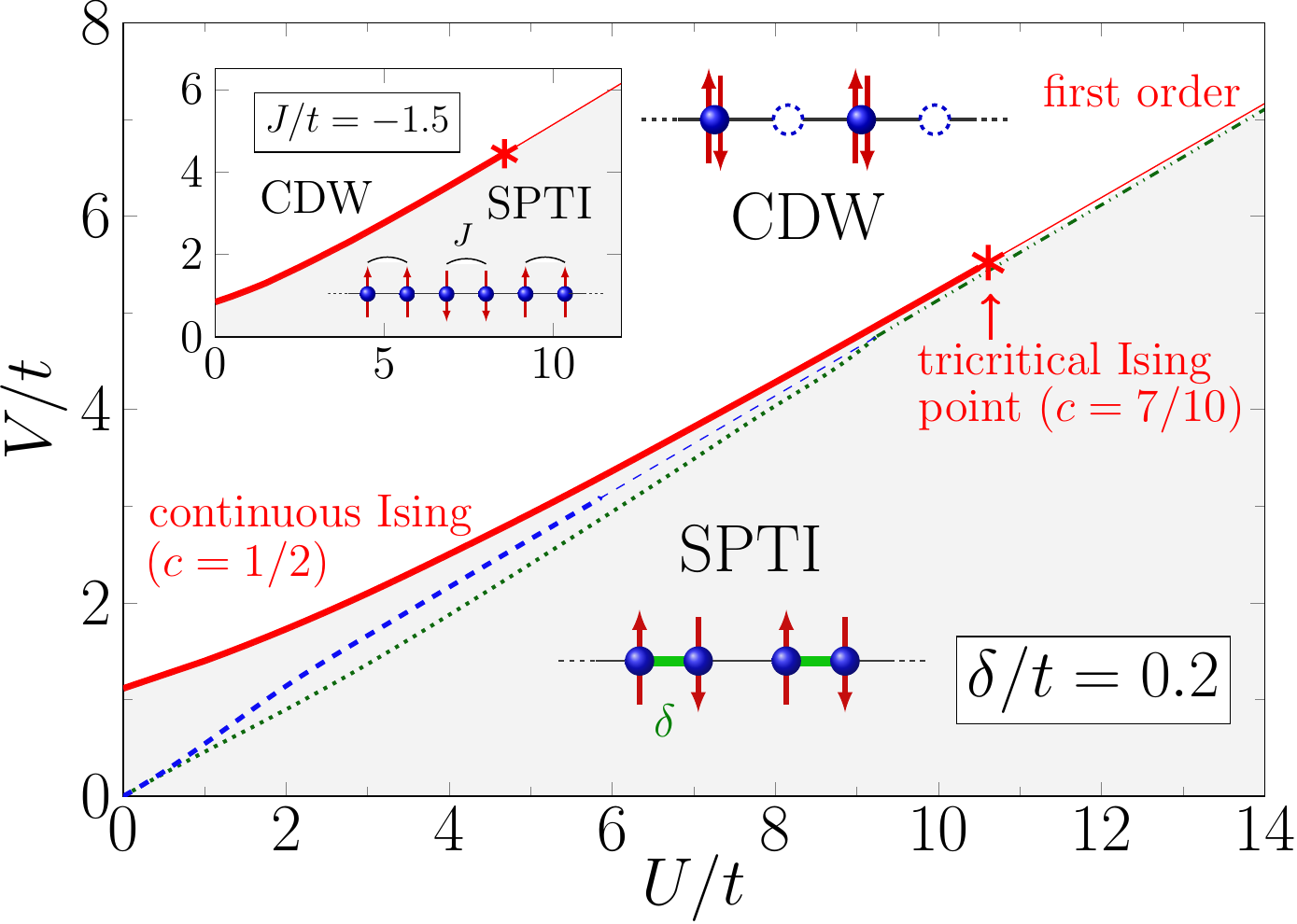}
\caption{\label{fig:symmetryprotection-ehmdelta} Phase diagram of the dimerized
extended Hubbard model [Eq.~(\ref{eq:EHMD})] from infinite-size
DMRG. The CDW--SPTI QPT is continuous with $c=1/2$ (first order) below (above) the tricritical 
Ising point where $c=7/10$~\cite{EELF16}. Dashed and dotted lines are the
CDW--BOW--SDW phase boundaries of the pure EHM
(cf. Fig.~\ref{fig:efkm-ehm}). Inset: phase diagram of the
extended Hubbard model with an alternating ferromagnetic Heisenberg interaction
[Eq.~(\ref{eq:EHMJ})] from infinite-size DMRG~\cite{TF04,GNT99,LEF15}. 
}
\end{figure}

We now explore the competition between traditional DW insulators
and SPT insulators (SPTIs). A prominent representative of an SPTI is the
Haldane insulator (HI) phase of the spin-$1$ Heisenberg chain~\cite{Hal83}. Recently, it has
been demonstrated that an SPT state also exists in the EHM with an additional
ferromagnetic spin interaction ($J<0$) on every other bond~\cite{LEF15},
\begin{align}\label{eq:EHMJ}
 \hat{H} = \hat{H}_{\textrm{EHM}}+J\sum_{i=1}^{L/2}\hat{\bm S}_{2i-1} \cdot \hat{\bm S}_{2i}\,.
\end{align}
Here, $\hat{\bm S}_i= \oh\sum_{\sigma\sigma^\prime}
      \hat{c}_{i\sigma^{\phantom{\prime}}}^\dagger 
      {\bm \sigma}_{\sigma\sigma^\prime}^{\phantom{\dagger}}
      \hat{c}_{i\sigma^\prime}^{\phantom{\dagger}}$.
Since the EHM behaves like a spin-$\oh$ chain for large $U/V$, the
Heisenberg term in Eq.~(\ref{eq:EHMJ}) promotes the formation of spin-$1$ moments 
from neighboring spins. The resulting effective antiferromagnetic spin-$1$
chain supports an HI phase with zero-energy edge excitations, similar to the spin-$1$ XXZ chain~\cite{PBTO12}. The SPTI
replaces the SDW and BOW phases of the EHM (Sec.~\ref{sec:extendedhubbard})
and reduces the extent of the CDW phase (see inset of
Fig.~\ref{fig:symmetryprotection-ehmdelta}). In the SPTI, the entanglement
spectrum shows a characteristic double-degeneracy of levels that is absent
in the topologically trivial CDW phase~\cite{LEF15}.
      
A similar scenario emerges in the context of carrier-lattice coupling. The
half-filled EHM with a staggered bond dimerization
$\delta$~\cite{TF04,BEG06,PhysRevB.86.205119,EELF16},
\begin{align}\label{eq:EHMD}
 \hat{H} = \hat{H}_{\textrm{EHM}}  
  -t \delta\sum_{i\sigma}(-1)^i (\hat{c}^\dagger_{i\sigma} \hat{c}_{i+1\sigma}^{\phantom{\dagger}}
 + {\rm H.c.})\,,
\end{align}
describes the formation of an SPT phase as a result of a Peierls
instability. The bond dimerization in Eq.~(\ref{eq:EHMD})
is equivalent to mean-field BOW order in the SSH model~(\ref{eq:ssh}). 

The phase diagram of the dimerized EHM~(\ref{eq:EHMD}) for $\delta=0.2$ was determined with
the infinite-size DMRG method and is shown in Fig.~\ref{fig:symmetryprotection-ehmdelta}~\cite{EELF16}.
The SDW and BOW phases of the pure EHM are entirely replaced by the SPTI. The
latter also has the lowest energy for $U=0$ and small $V/t$, which
confirms previous RG results~\cite{TF04} and leads to
a reduction of the CDW phase at weak couplings. The critical
line of the continuous Ising QPT terminates at a tricritical point,
above which the CDW--SPTI QPT becomes first order. The same holds for the
EHM with ferromagnetic spin exchange [Eq.~\eqref{eq:EHMJ}].  

\begin{figure}
\centering
\includegraphics[width=0.425\textwidth]{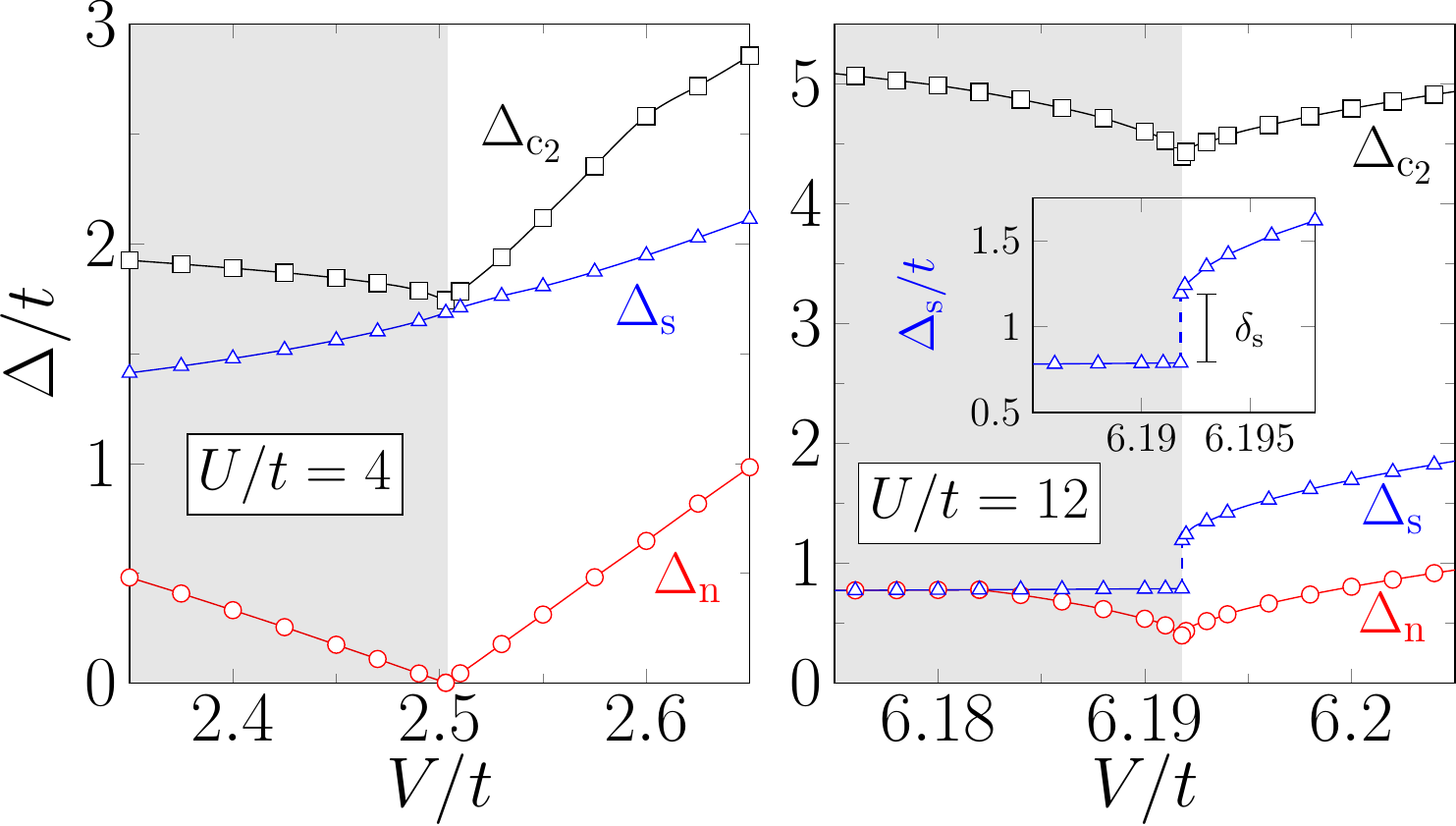}
\includegraphics[width=0.4675\textwidth]{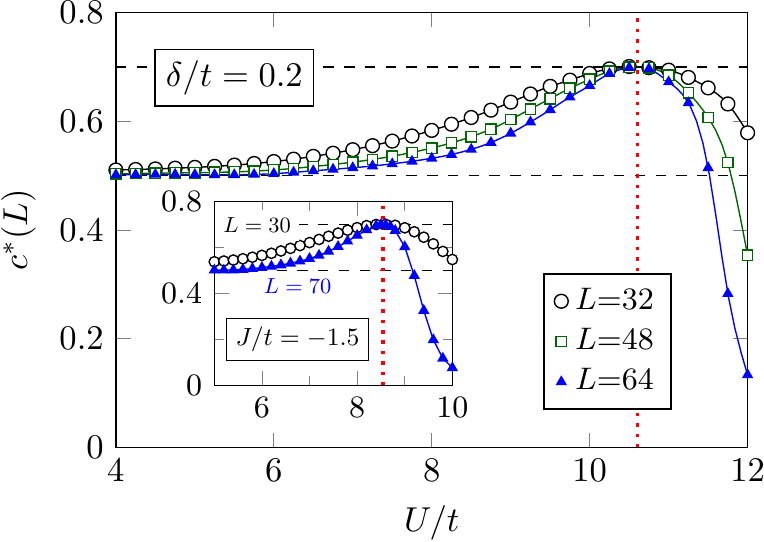}\hspace{1em}
\caption{\label{fig:symmetryprotection-gc} Top: DMRG charge ($\Delta_{\rm c_2}$), spin 
 ($\Delta_{\rm s}$), and neutral ($\Delta_{\rm n}$) gaps for the extended
 Hubbard model with bond dimerization [Eq.~(\ref{eq:EHMD})]. Here, $\delta/t=0.2$,  $U/t=4$
 (left) and $U/t=12$ (right).  The SPTI (CDW) phase is marked in gray (white). The spin gap exhibits
 a jump $\delta_{\rm s}\equiv \Delta_{\rm s} (V_{\rm c}^+) - \Delta_{\rm s}
 (V_{\rm c}^-)$  at $V_{c}/t$. Bottom: Central charge $c^\ast(L)$ along the
 CDW--SPTI transition line from DMRG calculations. The data indicate Ising universality ($c=1/2$) 
 for $U<U_\text{t}$ and, most notably, a tricritical Ising point 
 with $c=7/10$ at $U_\text{t}$ (red dotted line)~\cite{EELF16}. The inset shows results
 for the extended Hubbard model with an additional spin-spin interaction [Eq.~(\ref{eq:EHMJ})]~\cite{LEF15}.
}
\end{figure}

The various excitation gaps are shown in
Fig.~\ref{fig:symmetryprotection-gc}. For the pure EHM both
$\Delta_{\text{c}_2}$ and $\Delta_\text{n}$ vanish at the continuous CDW--BOW
QPT. For a nonzero dimerization $\delta$, the neutral gap closes whereas
$\Delta_{\text{c}_2}$ remains finite, indicating that the CDW--SPTI
QPT belongs to the Ising universality class. At strong coupling, 
all gaps remain finite across the QPT. The jump in the spin gap
$\Delta_\text{s}$ indicates a first-order transition. At very large $U$, the
low-lying excitations of Eq.~(\ref{eq:EHMD}) are related to the chargeless
singlet and triplet excitations of an effective spin-Peierls Hamiltonian.

At criticality, the central charge $c$ can
easily be extracted from the entanglement entropy~\cite{ELF14,EF15}. For
periodic boundary conditions, conformal field theory predicts the von Neumann entropy
to be $S_L(\ell)=\frac{c}{3}\ln \left[
  \frac{L}{\pi}\sin\left(\frac{\pi\ell}{L}\right) \right] +s_1$ where $s_1$
is a nonuniversal constant~\cite{CC04}. A finite-size estimate for the
central charge is then obtained via~\cite{Ni11} 
\begin{align}
 c^\ast(L) = \frac{3[S_L(L/2-2)-S_L(L/2)]}{\ln\{\cos[\pi/(L/2)]\}}\;,
\label{cstar}
\end{align}
taking the doubled  unit cell of the SPTI into account. The bottom panel of
Fig.~\ref{fig:symmetryprotection-gc} gives  $c^\ast(L)$, calculated  along
the PI-CDW QPT line by varying $U$ and $V$ simultaneously at fixed
dimerization. With increasing $U$ there is clear evidence for a crossover
from  $c^\ast(L)\simeq1/2$ to $c^\ast(L)\simeq7/10$, signaling Ising
tricriticality. A bosonization-based field-theory analysis of the power-law
(exponential) decay of the CDW, SDW, and BOW correlations confirms the
universality class of the tricritical Ising model~\cite{EELF16}.

\subsection{Extended anyon-Hubbard model}\label{sec:eahm}

\begin{figure}
  \centering
  \includegraphics[width=0.4\textwidth,trim=0 1.5em 0 0,clip]{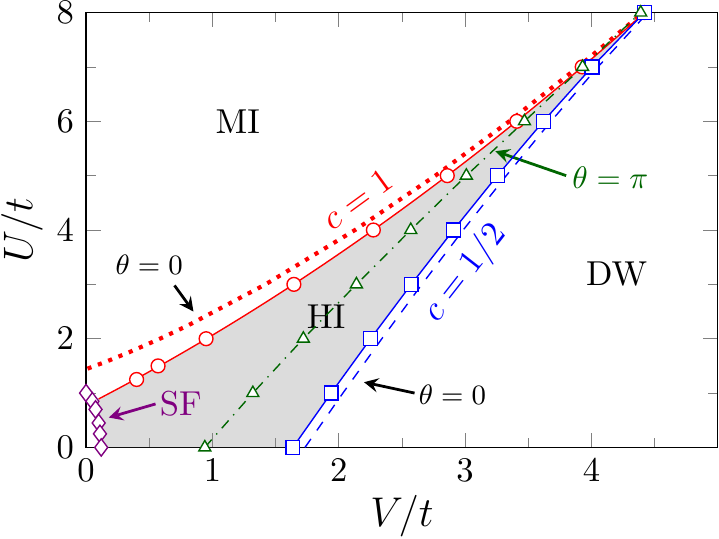}\\
  \includegraphics[width=0.4\textwidth,trim=0 1.5em 0 0,clip]{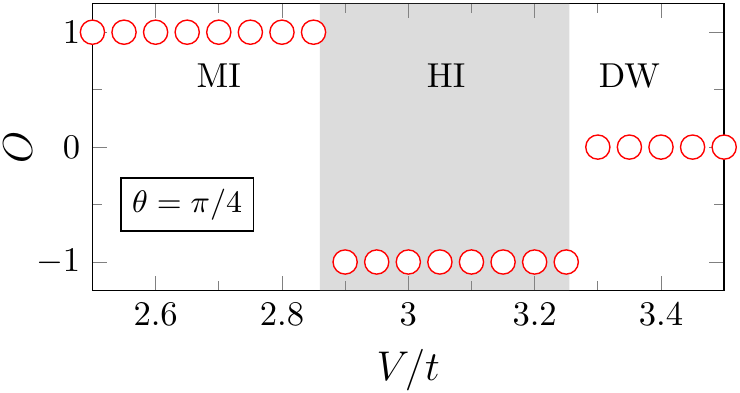}\\
  \includegraphics[width=0.4\textwidth]{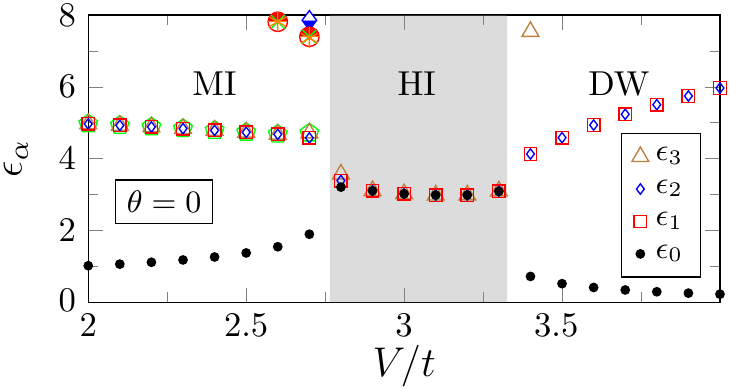}
  \caption{\label{fig:exhm} Top: Phase diagram of the extended anyon-Hubbard
    model~(\ref{eq:eahm}) with $n=1$, $n_\text{p}=2$, $\theta=\pi/4$~\cite{LEF17}. 
    Dotted (dashed) lines mark the MI--HI (HI--DW) QPT
    in the extended bosonic Hubbard model ($\theta=0$)~\cite{ELF14}.  The dashed-dotted line with
    triangles up marks the first-order MI--DW QPT for
    $\theta=\pi$. Middle: Order parameter $O$ (see text) for the EAHM with $U/t=5$. Bottom: DMRG data for the entanglement
    spectrum $\xi_\alpha$ of the extended bosonic Hubbard model with $U/t=5$.}
\end{figure}

Ultracold atomic gases in optical lattices provide the possibility to 
study not only fermions or bosons but also anyons. Exchanges of the latter 
result in a phase factor $e^{\mathrm{i}\theta}$ in the many-body wave
function. The statistical parameter $\theta$ can take on any value between
$0$ and $\pi$, so that anyons interpolate between bosons and
fermions~\cite{LM77,Wi82}.  With Haldane's generalized Pauli principle~\cite{Ha91}, the anyon concept becomes important
also in 1D systems. A fascinating question is if the HI phase observed, e.g.,
in the extended Bose-Hubbard model (EBHM)~\cite{TBA06,EF15} also exists in the extended anyon-Hubbard model (EAHM).

After a fractional Jordan-Wigner transformation of the anyon operators,
$\hat{a}_i\mapsto \hat{b}_i e^{\mathrm{i}\theta\sum_{l=1}^{i-1}\hat{n}_l}$
\cite{KLMR11}, the Hamiltonian of the EAHM takes the form~\cite{LEF17} 
\begin{align}
 \hat{H}_{\rm EAHM}=&-t\sum_i
  (\hat{b}_i^{\dagger} \hat{b}_{i+1}^{\phantom{\dagger}} e^{\mathrm{i}\theta \on_i}
   +e^{-\mathrm{i}\theta \on_i}  \hat{b}_{i+1}^{\dagger} \hat{b}_j^{\phantom{\dagger}})\nonumber\\
   &+U\sum_{i} \hat{n}_i (\hat{n}_i-1)/2 +V\sum_{i} \hat{n}_i \hat{n}_{i+1}\,,
  \label{eq:eahm}
\end{align} 
where $\hat{b}_i^\dagger$ ($\hat{b}_i^{\phantom{\dagger}}$) is a bosonic creation (annihilation) operator, and
$\hat{n}_i=\hat{b}_i^\dagger \hat{b}_i^{\phantom{\dagger}}=\hat{a}_i^\dagger \hat{a}_i^{\phantom{\dagger}}$.
A boson hopping from site $i+1$ to site $i$ acquires an occupation-dependent
phase. Note that anyons on the same site behave as ordinary bosons. Anyons
with $\theta=\pi$ represent so-called ``pseudofermions'', namely, they are
fermions offsite but bosons onsite. If the maximum number of particles per site is restricted to
$n_\text{p}=2$, the EBHM---the  $\theta \to 0$ limit of the
EAHM~\eqref{eq:eahm}---maps to an effective XXZ spin-$1$ chain~\cite{BTGA08}.
  
The phase boundaries of the EAHM (EBHM) with $\theta =\pi/4$ ($\theta =0$)
and $n_\text{p}=2$ are shown in the top panel of Fig.~\ref{fig:exhm}. Most notably, the HI---located between MI
and DW insulating phases in the EBHM---survives for any finite
fractional phase, i.e., in the anyonic case~\cite{LEF17}. Likewise, the
superfluid (SF) appears for very weak coupling. The critical values for the MI--HI
QPT (squares) and the HI--DW QPT (circles) were determined from a divergence of the
correlation length $\xi_\chi$ with increasing DMRG bond-dimension $\chi$;
the model becomes critical with central charge $c=1$ and $c=1/2$,
respectively. 

The HI may naively be expected to disappear in the EAHM with $\theta>0$
which has neither time reversal ($\hat{\cal T}$) nor inversion ($\hat{\cal
  I}$) symmetry.  However, it has been shown that there exists a 
nontrivial topological phase protected by the combination of  $\hat{\cal R}^z =
e^{\mathrm{i}\pi\sum_j(\on_i-1)}$ and  $\hat{\cal K} =e^{\mathrm{i}\theta\sum_i
  \on_i(\on_i-1)/2}$ $\hat{\cal I}\,\hat{\cal T}$~\cite{LEF17}. A nonlocal order
parameter ${O}$ can be constructed that discriminates between states that are
symmetric under both $\hat{\cal K}$ {\it and} $\hat{{\cal R}^z}$ and states
that are not. The middle panel of Fig.~\ref{fig:exhm}
demonstrates that ${O}$ can be used to distinguish the topologically
trivial MI and DW phases (${O}=1$) from the topologically nontrivial HI
(${O}=-1$). 

Valuable information about topological phases is also provided by the
entanglement spectrum $\{\xi_\alpha\}$~\cite{LH08}. The concept of
entanglement is inherent in any DMRG algorithm based on matrix-product states. Dividing the system into two subsystems,
$\xi_\alpha=-2\ln\lambda_\alpha$ is determined by the singular values
$\lambda_\alpha$ of the reduced density matrix~\cite{LEF15}. The lower panel
of Fig.~\ref{fig:exhm} shows the entanglement spectrum of the EBHM with
$U/t=5$.  In the HI phase the entanglement spectrum is expected to be at
least four-fold degenerate,  reflecting the broken
$\mathbb{Z}_2\times\mathbb{Z}_2$ symmetry.  This is clearly seen in the
HI phase.  By contrast, in the trivial MI and DW phases, the lowest
entanglement level is always nondegenerate. 

\section{Conclusions}\label{sec:conclusions}

The 1D correlated quantum systems reviewed here exhibit a
remarkably rich variety of physical properties that can be studied and
understood in particular by powerful numerical methods. For the case of
half-filled bands considered, metallic phases are either spinless TLLs or
spinful LELs of repulsive nature, \ie, with dominant CDW or BOW correlations. The insulating
phases fall into three categories: (i) long-range ordered with a spontaneously broken Ising
symmetry (BOW, CDW), (ii) critical with no symmetry breaking (SDW, EI), (iii)
topologically nontrivial with short-range entanglement (HI). While the existence
of these phases and the phase transitions can in principle be inferred from
the low-energy field theory, the details for a given microscopic
model typically require numerical solutions. In particular, mean-field,
variational or even bosonization/RG approaches are in general not sufficient,
especially for problems with retarded interactions.

Despite the significant advances reviewed here, 1D correlated quantum systems remain an active, rewarding and
challenging topic of condensed matter physics. Even with the physics of the most
fundamental models now unraveled, there remain many future problems of
importance in relation to experiment. The list of topics includes the effect
of Jahn-Teller coupling at finite band filling, competing long-range
interactions, thermodynamics, time-dependent or nonequilibrium
phenomena, as well as
the coupling to a substrate or other chains. 

\section*{Acknowledgments}
MH was supported by the German Research Foundation (DFG) through SFB 1170
ToCoTronics and FOR 1807 ({\it Advanced Computational Methods for Strongly Correlated Quantum Systems}), HF by SFB 652 und the priority programme 1648 ({\it Software for Exascale
  Computing}). We are grateful to A. Alvermann, F. F. Assaad, D. M. Edwards,
S. Ejima, F. H. L. Essler, G. Hager, E. Jeckelmann, F. Lange, M. Weber, A. Wei{\ss}e, 
and G. Wellein for fruitful collaboration. We also thank S. Ejima for preparing
some of the figures.

\section*{Author contributions}
Both authors contributed equally to  the preparation of the manuscript and approved
it in its final form.


\end{document}